\documentclass[aps,prb,reprint,amsmath,amssymb,superscriptaddress]{revtex4-2}

\usepackage[T1]{fontenc}
\usepackage[utf8]{inputenc}
\usepackage{color}
\usepackage{babel}
\usepackage{verbatim}
\usepackage{float}
\usepackage{lmodern}
\usepackage{amsmath,amssymb,dsfont,amstext,amsfonts}
\usepackage{bbold}
\usepackage{mathrsfs}
\usepackage{mathtools}

\usepackage{graphicx}
\usepackage[unicode=true,pdfusetitle,
 bookmarks=true,bookmarksnumbered=false,bookmarksopen=false,
 breaklinks=true,pdfborder={0 0 1},backref=false,colorlinks=true]
 {hyperref}
\usepackage{svg}

\newcommand{\ie}{{\it i.e.}\;}
\newcommand{\eg}{{\it e.g.}\;}

\makeatletter

\usepackage{graphicx}
\usepackage{epstopdf}
\usepackage{epsfig}

\usepackage[normalem]{ulem}

\definecolor{AB}{rgb}{0.2,0.1,0}
\definecolor{GY}{rgb}{0.1,0,0}

\makeatother

\begin{document}
\title{Landau Levels of the Euler Class Topology}
\begin{abstract}

Two-dimensional systems with $C_{2}\mathcal{T}$ ($P\mathcal{T}$) symmetry exhibit the Euler class topology $E\in\mathbb{Z}$ in each two-band subspace realizing a fragile topology beyond the symmetry indicators. By systematically studying the energy levels of Euler insulating phases in the presence of an external magnetic field, we reveal the robust gaplessness of the Hofstadter butterfly spectrum in the flat-band limit, while for the dispersive bands the gapping of the Landau levels is controlled by a hidden symmetry.
We also find that the Euler class $E$ of a two-band subspace gives a lower bound for the Chern numbers of the magnetic subgaps. Our study provides new fundamental insights into the fragile topology of flat-band systems going beyond the special case of $E=1$ as {\it e.g.}\; in twisted bilayer graphene, thus opening the way to a very rich, still mainly unexplored, topological landscape with higher Euler classes.

\end{abstract}

    \author{Yifei Guan$^1$}\thanks{Contributed equally.  Correspondence to \href{mailto:yifei.guan@epfl.ch}{yifei.guan@epfl.ch} and \href{mailto:adrien.bouhon@su.se}{adrien.bouhon@su.se}.}

\author{Adrien Bouhon$^2$}\thanks{Contributed equally.  Correspondence to \href{mailto:yifei.guan@epfl.ch}{yifei.guan@epfl.ch} and \href{mailto:adrien.bouhon@su.se}{adrien.bouhon@su.se}.}

\author{Oleg V. Yazyev$^{1,3}$}

\affiliation{\vspace*{0.4cm}$^{1}$Institute of Physics, \'{E}cole Polytechnique F\'{e}d\'{e}rale de Lausanne (EPFL), CH-1015 Lausanne, Switzerland}
\affiliation{$^{2}$Nordita, Stockholm University and KTH Royal Institute of Technology, Hannes Alfv{\'e}ns v{\"a}g 12, SE-106 91 Stockholm, Sweden}
\affiliation{$^{3}$National Centre for Computational Design and Discovery of Novel Materials MARVEL, \'{E}cole Polytechnique F\'{e}d\'{e}rale de Lausanne (EPFL), CH-1015 Lausanne, Switzerland}

\date{\today}

\maketitle

\section{Introduction}

Since the discovery of the integer quantum Hall effect (QHE) \cite{KlitzingQHE,von1986quantized} the concept of topology has played an increasing role in condensed matter physics \cite{Laughlin81,thouless1982quantized,Stone92,Thouless98,Simon1983_homotopy,avron1989,Hatsugai_BBC}. 
The prediction of the quantum spin-Hall effect \cite{KaneQSHE,KaneMeleZ2,BernevigQSHE,Bernevig1757} and the three-dimensional topological insulators (TI) \cite{Fu3D,Chen178,SCZhang2011} protected by time-reversal symmetry have then opened the way to the realization of many novel electronic states and has attracted much attention to the topological aspects of electronic band structures. 
The role of symmetries has proven essential for the tenfold classification of topological phases of matter \cite{Schnyder08,Kitaev09_AIP} and its extension to crystalline symmetries \cite{FuKane_inversion,TurnerInversion,Fu_crystalline,Bernevig2012_point_group,Slager2013,RyuMirror,Schnyder2014Mirror,Shiozaki14,Wi3,SchnyderClass,Cornfeld2019}. 
This has recently culminated in systematic classification schemes that address the global band structure topology \cite{Watanabe_connectivity} in terms of irreducible representation combinatorics \cite{Slager_combinatorics,Bouhon_global,ShiozakiSatoGomiK}, symmetry-based indicators \cite{Po2017,Khalaf2018a}, topological quantum chemistry \cite{Bradlyn2017,HolAlex_Bloch_Oscillations}, and real-space topological crystals \cite{Songeaax2007,SongRealSpace2020,PhysRevX.8.011040,shiozaki2018spectral}. 
The discrepancy between the stable symmetry indicators and the topology of split elementary band representations has then led to the definition of crystalline fragile topology for few-band subspaces \cite{po2018fragile,Bouhon2018,BJY_nielsen,SongMonoids,Peri2019,Song794}.

In its most intriguing form, fragile topology arises without symmetry indicators and is protected by an anti-unitary symmetry that squares to $+1$ and leaves the momentum invariant, e.g.~$P\mathcal{T}$ symmetry in spinless systems, or $C_2\mathcal{T}$ symmetry in two-dimensional spinless and spinful materials, in which case it is called the Euler class topology \cite{BJY_linking,Bouhon2019c,BouhonGEO2020,Zhao_PT}. 
Two-dimensional Euler insulating phases have been found to exhibit very rich physics, ranging from the non-Abelian braiding of nodal points \cite{Wu1273,BJY_nielsen} in electronic band structures \cite{Bouhon2019c,chen2021manipulation}, in acoustic metamaterials \cite{jiang2021observation} and in the phonon band structures of silicates \cite{peng2021nonabelian} and Al$_2$O$_3$ \cite{peng2022multigap}, where it also explains the stability of the Goldstone modes degeneracy at $\Gamma$ \cite{park2021acoustic,lange2022continuum}. Furthermore, the Euler class topology has been found at the origin of Hopf linking signatures in quenched optical lattices \cite{UnalQuench}, and in the topology in magic-angle twisted bilayer graphene (TBG) \cite{Po_TBG,SongMonoids,PhysRevB.99.155415}. 

The unveiling of further robust physical signatures for the Euler class topology is very timely. 
Recently, the effect of an external magnetic field on effective models of the moir{\'e} flat bands in TBG has been reported \cite{Lian2020b}, as well as in other twisted bilayer systems \cite{Lian2020}. By facilitating much higher magnetic flux per unit cell, moir\'{e} super-lattices represent a great venue for the measurement of the Hofstadter butterfly spectrum \cite{dean2013hofstadter,Dana1985e} as shown in \cite{cao2018correlated,Wu2020}. 
Reversely, the effect of different band structure topologies on the Hofstadter spectrum has been shown to lead to rich distinctive features \cite{Herzog-Arbeitman2020a}.

In this work, we study the effect of the Euler class topology on the Hofstadter spectrum of two-dimensional systems. 
We reveal qualitative signatures of the Euler class in the flat-band limit and more general non-degenerate and dispersive (non-flat) band structures. 
In particular, we provide the first systematic study of balanced and imbalanced Euler topological phases, which are characterized by equal, and, respectively, distinct Euler classes below and above the energy gap. 
While the flat-band limit exhibits a robust gapless Hofstadter spectrum, we unveil a hidden symmetry that controls the gaplessness of the Hofstadter spectrum of the dispersive balanced Euler insulators. 
We furthermore show that the Hofstadter spectrum of the imbalanced Euler phases is generically gapless.

\section{Euler class topology}

The $C_{2z}\mathcal{T}$ symmetry of a two-dimensional system has $[C_{2z}\mathcal{T}]^2=+1$ and leaves the momentum of the Bloch states invariant within the two-dimensional Brillouin zone. This guarantees the existence of a basis with a real and symmetric Bloch Hamiltonian, $H\rightarrow \widetilde{H} = \widetilde{H}^T \in \mathbb{R}^N \times \mathbb{R}^N$ \cite{Bouhon2019c}. We are here excluding non-orientable phases characterized by $\pi$-Berry phases along some non-contractible loops of the Brillouin zone \cite{BJY_linking,BouhonGEO2020}.
The Euler class $E\in\mathbb{Z}$ of real oriented rank-2 vector bundles \cite{Hatcher_2} then characterizes the two-dimensional topology of every (orientable) two-band vector subspace $V^{\alpha}$ of the band structure, which we label by $\alpha=I,II,\dots$, {\it i.e.}~$V^{\alpha} = \langle u^{\alpha}_a,u^{\alpha}_b \rangle_{\mathbb{R}^2} $ is the vector space spanned by the eigenvectors $\{u^{\alpha}_{n}\}_{n=a,b}$ corresponding to the eigenvalues $\{\epsilon^{\alpha}_n\}_{n=a,b}$ obtained from the spectral decomposition $\widetilde{H} u^{\alpha}_n = \epsilon^{\alpha}_n u^{\alpha}_n$, where we assume the energy ordering $\epsilon^{\alpha}_n < \epsilon^{\alpha+I}_n$ ($n=a,b$). Then, the topology for any group of bands with more than two bands is reduced to the second Stiefel-Whitney class $w_2 = E \, \mathrm{mod} \, 2  \in \mathbb{Z}_2$ \cite{BJY_linking,BJY_nielsen,Bouhon2019c,BouhonGEO2020,Zhao_PT}, \ie there is a $\mathbb{Z}\rightarrow \mathbb{Z}_2$ reduction specific to fragile topology. In particular, any two-band subspace with an even Euler class is trivialized when a third (trivial) band is added to the band-subspace (see Fig.\;3 in \cite{Supple}). The Euler class of the $\alpha$-th two-band subspace is computed through the integral \cite{BJY_linking,BJY_nielsen,Xie2020_superfluid,Bouhon2019c,Zhao_PT} 
\begin{equation}
    E_{\alpha} = \dfrac{1}{2\pi}\int_{\mathrm{BZ}} dk_1 dk_2 \, \mathrm{Eu}_{\alpha},
\end{equation}
over BZ=$[-\pi,\pi)^2$, the Brillouin zone of the two-dimensional lattice, with the integrand given by the Euler curvature
\begin{equation}
\label{eq_Euler_class}
\begin{aligned}
    \mathrm{Eu}_{\alpha} &= \mathrm{Pf} F[(u^{\alpha}_a~u^{\alpha}_b)],\\
    &= (\partial_{k_1} u^{\alpha}_a)^T  \cdot (\partial_{k_2} u^{\alpha}_b) - (\partial_{k_2} u^{\alpha}_a)^T  \cdot (\partial_{k_1} u^{\alpha}_b),
\end{aligned}
\end{equation}
here defined as the Pfaffian of the two-state Berry curvature  
\begin{equation}
\begin{aligned}
    F[(u^{\alpha}_a~u^{\alpha}_b)] &=
    F[\vert \boldsymbol{u}_{\alpha} \rangle],\\ 
    &= -\mathrm{i}\,\left(\langle \partial_{k_1} \boldsymbol{u}_{\alpha} \vert \partial_{k_2} \boldsymbol{u}_{\alpha} \rangle - \langle \partial_{k_2} \boldsymbol{u}_{\alpha} \vert \partial_{k_1} \boldsymbol{u}_{\alpha} \rangle\right),
\end{aligned}
\end{equation}
with the matrix of two column eigenvectors $\vert \boldsymbol{u}_{\alpha} \rangle = (u^{\alpha}_a~u^{\alpha}_b)$. Alternatively, the Euler class can be obtained as the winding of the two-band Wilson loop \cite{Bouhon2018,BJY_linking,Xie2020_superfluid,Bouhon2019c,BJY_nielsen}, see Fig.\;\ref{fig:bandwcc}. Interestingly, Eq.\;(\ref{eq_Euler_class}) motivates yet another way to compute the Euler class. Defining the \textit{Chern basis}
\begin{equation}
\label{eq_chern_basis}
\begin{aligned}
v^{\alpha}_+ & = (u^{\alpha}_a+\mathrm{i} u^{\alpha}_b)/\sqrt{2},\\
v^{\alpha}_- & = (u^{\alpha}_a-\mathrm{i} u^{\alpha}_b)/\sqrt{2},
\end{aligned}
\end{equation}
and writing the one state Berry curvature $F[v^{\alpha}_+] = (\partial_{k_1} v^{\alpha}_-)^T \cdot (\partial_{k_2} v^{\alpha}_+) - (\partial_{k_2} v^{\alpha}_-)^T \cdot  (\partial_{k_1} v^{\alpha}_+)$,
we readily find $\mathrm{Eu}_{\alpha} = F[v^{\alpha}_+] $, from which we obtain \cite{Bouhon2019c} the Euler class as a one-band Chern number 
\begin{equation}
E_{\alpha}  =  \dfrac{1}{2\pi} \int_{\mathrm{BZ}} dk_1dk_2\,F[v^{\alpha}_+]  =  C  .
\end{equation}
In the limit of degenerate bands ($\epsilon^{\alpha}_a=\epsilon^{\alpha}_b$), the Chern basis becomes an eigenbasis of $\widetilde{H}$. 
This plays an important role in the flat-band limit discussed below. An essential observable associated with the Euler class $E_{\alpha}$ is the number $2 \vert E_{\alpha}\vert $ of \textit{stable} nodal points hosted by the $\alpha$(=$I,II$)-th two-band subspace (\ie the nodes cannot be annihilated as long as the energy gaps above and below the two-band subspace remain open), see {\it e.g.}~the four stable nodes in each two-band subspace of Fig.~\ref{fig:bandwcc}(a) for the Euler phase with $E_I=E_{II}=2$. These nodes cannot be annihilated as long as the energy gaps above and below the two-band subspace remain open \cite{BJY_nielsen,Bouhon2019c,BouhonGEO2020}. This must be contrasted for instance with the two nodes of graphene that can be annihilated upon breaking the $C_6$ crystal symmetry while preserving $C_2\mathcal{T}$ symmetry.

\section{Four-band real symmetric Hamiltonian}

In the following, we consider a four-orbital system that is insulating at half-filling $\nu = N_{\mathrm{occ}}/N_{\mathrm{orb}} = 1/2$, with $N_{\mathrm{orb}}$ the total number of orbitals (\ie either four spinless orbitals, or two spin-$1/2$ pairs) and $N_{\mathrm{occ}}$ the number of bands below the energy gap. The most general four-band real symmetric Bloch Hamiltonian is spanned by nine real independent terms, {\it i.e.}
\begin{equation}
\label{eq_H_general}
    \widetilde{H} = \sum_{i,j=0,x,y,z} h_{ij} \Gamma_{ij},
\end{equation}
for $\Gamma_{ij} = \sigma_i\otimes \sigma_j$ and $i,j=0,x,y,z$ with $\sigma_{x,y,z}$ the Pauli matrices and $\sigma_0 = \mathbb{1}$, under the constraint that only the terms with $\Im\Gamma_{ij} = \mathbb{0}$ are kept. Thus, the most general Bloch Hamiltonian is parametrized by only ten parameters, 
\begin{equation}
\label{eq_ten_terms}
    \{h_{00},h_{0x},h_{0z},h_{x0},h_{xx},h_{xz},h_{yy},h_{z0},h_{zx},h_{zz}\}\in \mathbb{R}.  
\end{equation}
In the following, we,discard the term $h_{00}$ since it does not affect the topology. 

We first consider the Hofstadter butterfly in the limit of flat bands. The flat-band limit of the Euler insulating phases implies the two-by-two degeneracy of the bands since each two-band subspace with a non-zero Euler class hosts stable nodal points, as we have seen above. The most general four-band Bloch Hamiltonian (real and symmetric) with a gapped and flat spectrum, \ie we set the eigenvalues to $(\epsilon_1,\epsilon_2,\epsilon_3,\epsilon_4)=(-1,-1,1,1)$, takes the form (see Appendix A) 
\begin{equation}
\label{eq_H_imb}
\begin{aligned}
    \widetilde{H}[\boldsymbol{n},\boldsymbol{n}'] &= n_1' (
        -n_1 \Gamma_{zz} + n_2 \Gamma_{zx} + n_3 \Gamma_{x0}
    )& \\
     &- n_2' (
        +n_1 \Gamma_{xz} - n_2 \Gamma_{xx} + n_3 \Gamma_{z0}
    )& \\
    &+ n_3' (
       + n_1 \Gamma_{0x} + n_2 \Gamma_{0z} - n_3 \Gamma_{yy}
    )&,
\end{aligned}
\end{equation}
which depends on two unit vectors 
\begin{equation}
\begin{aligned}
    \boldsymbol{n}^{(\prime)} &= (n_1^{(\prime)},n_2^{(\prime)},n_3^{(\prime)}) \\
    &= (\cos\phi^{(\prime)} \sin\theta^{(\prime)},\sin\phi^{(\prime)} \sin\theta^{(\prime)},\cos\theta^{(\prime)}).
\end{aligned}
\end{equation}
The Bloch Hamiltonian Eq.\;(\ref{eq_H_imb}) thus defines a mapping from the Brillouin zone onto two unit spheres, $(k_1,k_2) \mapsto (\boldsymbol{n},\boldsymbol{n}') \in (\mathbb{S}^2, \mathbb{S}^{2'})$, the Fourier transform of which defines a tight-binding model \cite{BouhonGEO2020}. The topology of Eq.~(\ref{eq_H_imb}) is then determined by two skyrmion numbers, 
\begin{equation}
    q=W[\boldsymbol{n}]~\text{and}~ q'=W[\boldsymbol{n}'],  
\end{equation}
computing the winding of the unit vectors through
\begin{equation}
W[\boldsymbol{n}^{(\prime)}]  =  \dfrac{1}{4\pi} \int_{\mathrm{BZ}} dk_1dk_2 \,\boldsymbol{n}^{(\prime)}\cdot (\partial_{k_1} \boldsymbol{n}^{(\prime)}\times \partial_{k_2} \boldsymbol{n}^{(\prime)})  \in \mathbb{Z}.
\end{equation}
The Euler classes of the two two-band subspaces are then readily obtained from the Skyrmion numbers \cite{BouhonGEO2020} (see Appendix A)
\begin{equation}
    E_{I} =  q-q' ,\quad E_{II} =  q+q'.
\end{equation}
We importantly note that the sign of the Euler classes can be flipped in pair, \ie $(E_I,E_{II})\rightarrow (-E_I,-E_{II})$, under an adiabatic transformation of the Hamiltonian which is obtained from the nontrivial action of the generator of $\pi_1[\mathsf{Gr}_{2,4}^{\mathbb{R}}]=\mathbb{Z}_2$ on $\pi_2[\mathsf{Gr}_{2,4}^{\mathbb{R}}] = \mathbb{Z}^2$ \cite{BouhonGEO2020}. 

\begin{figure}[t!]
	\centering
	\includegraphics[width=8.6cm]{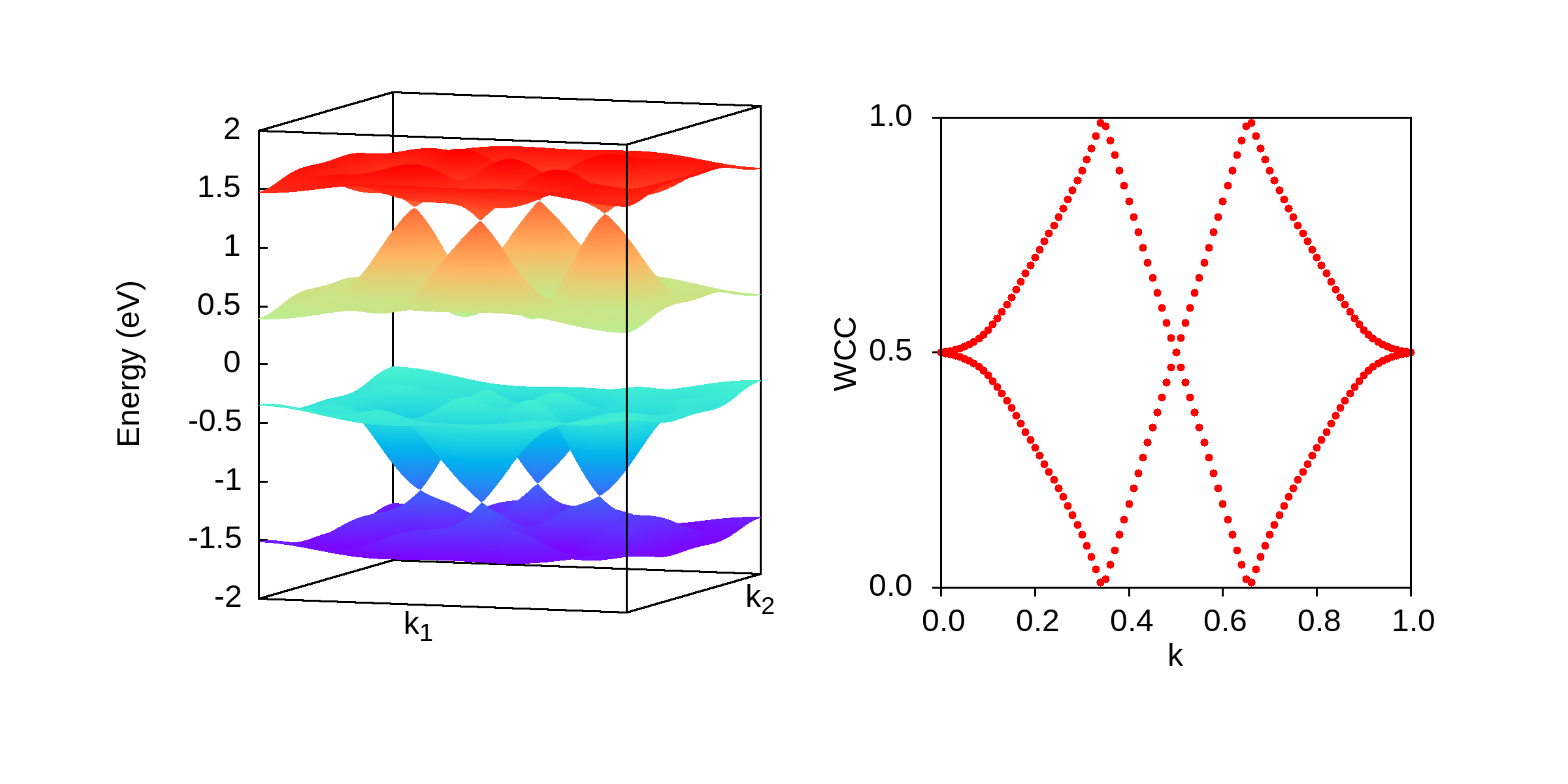}
	\caption{(a) Band structure of the tight-binding model of the balanced phase with the Euler class $E_{I}=E_{II}=2$. Four nodal points are present in each two-band subspace as a manifestation of the nontrivial Euler topology. (b) The value of the Euler class can be computed through the winding number, here $2$, of the two-band Wilson loop.
	}
	\label{fig:bandwcc}
\end{figure}
In the following we distinguish between two classes of phases, the \textit{balanced} phases for which $\vert E_I \vert = \vert E_{II}\vert $, and the \textit{imbalanced} phases with $\vert E_{I} \vert \neq \vert E_{II} \vert$. The balanced phases are readily obtained by setting one skyrmion number to zero, {\it e.g.}~fixing $\boldsymbol{n}'=(0,1,0)$ Eq.\;(\ref{eq_H_imb}) gives 
\begin{equation}
\label{eq_bal_H}
\begin{aligned}
    \widetilde{H}_{\mathrm{bal}}[\boldsymbol{n}] &= \widetilde{H}[\boldsymbol{n},(0,1,0)],\\ 
    &= - n_1 \Gamma_{xz} + n_2 \Gamma_{xx} - n_3 \Gamma_{z0},
\end{aligned}
\end{equation}
which is characterized by $q'=0$ and $E_I = E_{II} = q $. Moreover, it can be shown that every pair of balanced phases $(E_I,E_{II}) = (\bar{q},\bar{q})$, i.e. setting $q=\bar{q}$ and $q'=0$, and $(E_I,E_{II}) = (-\bar{q},\bar{q})$, i.e. setting $q=0$ and $q'=\bar{q}$, are homotopy equivalent \cite{bouhon2022multigap}. The imbalanced phases are then realized when both skyrmion numbers $q$ and $q'$ are nonzero, such that the nine terms in Eq.~(\ref{eq_H_imb}) are nonzero. %
Limiting ourselves to $ E_{I}+E_{II} \leq 4$ and $E_I,E_{II}\geq 0$, we discuss the balanced phases for $(E_I,E_{II})=(1,1), (2,2)$,
and the imbalanced phases for $(E_I,E_{II}) = (0,2), (1,3)$.
Fig.\;\ref{fig:bandwcc} presents the band structure and the Wilson loop of the balanced phase ($E_{I}$,$E_{II}$)=(2,2), and the other phases are shown in the supplementary material \cite{Supple}. Since we do not find any qualitative difference in the Hofstadter spectrum between the phases $(E_I,E_{II})\sim (-E_I,-E_{II})$ and $(E_I,-E_{II}) \sim (E_I,-E_{II})$, it is enough to show the results for $E_I,E_{II}\geq 0$.

\section{Hofstadter Spectrum}

The effect of an external magnetic field $\boldsymbol{B} = \boldsymbol{\nabla}\times \boldsymbol{A}$ is most conveniently introduced through the Peierls substitution $t_{ij} \rightarrow \widetilde{t}_{ij} = t_{ij} \exp(\mathrm{i} \phi_{ij})$ with $\phi_{ij} \propto \int_{\boldsymbol{R}_i}^{\boldsymbol{R}_j} \boldsymbol{A} \cdot d\boldsymbol{r}$ \cite{Graf1995}. 
Restricting to a rational magnetic flux, {\it i.e.}~$\phi = \int_{\mathrm{u.c.}} B d^2\boldsymbol{r} = 2 (r/s) \phi_0$ with $r$ and $s$ coprime integers ($\phi_0 = h/e$ is the magnetic flux quantum), the magnetic tight-binding Hamiltonian acquires a reduced periodicity with a magnetic unit cell \cite{Herzog-Arbeitman2020a} that is $s$ times as large as the non-magnetic one \cite{Supple}. 
It follows that the rotation $C_{2z}$ acts as a nontrivial permutation of the $s$ sub-lattice sites of the magnetic unit cell, leading to the breaking of $C_{2z}\mathcal{T}$ symmetry. We emphasize that while $C_{2z}$ is not necessarily a symmetry of the Hamiltonian, $C_{2z}\mathcal{T}$ alone imposes the $C_{2z}$-symmetric spatial configuration of the atomic orbitals since $\mathcal{T}$ does not affect the position operator. In other words, any orbital $\varphi_{\beta}$ located away from a $C_{2z}$ center, say $\boldsymbol{r}_{\beta}$, must have a $C_{2z}$ partner located at $C_{2z}\boldsymbol{r}_{\beta}$. As a consequence of the breaking of $C_{2z}\mathcal{T}$ by the external magnetic field, the $2 E_{\alpha}$ nodal points of each two-band subspace become gapped leading to magnetic Chern bands, see Section \ref{sec_dis_band}. 

\begin{figure}
	\centering
	\includegraphics[width=8cm]{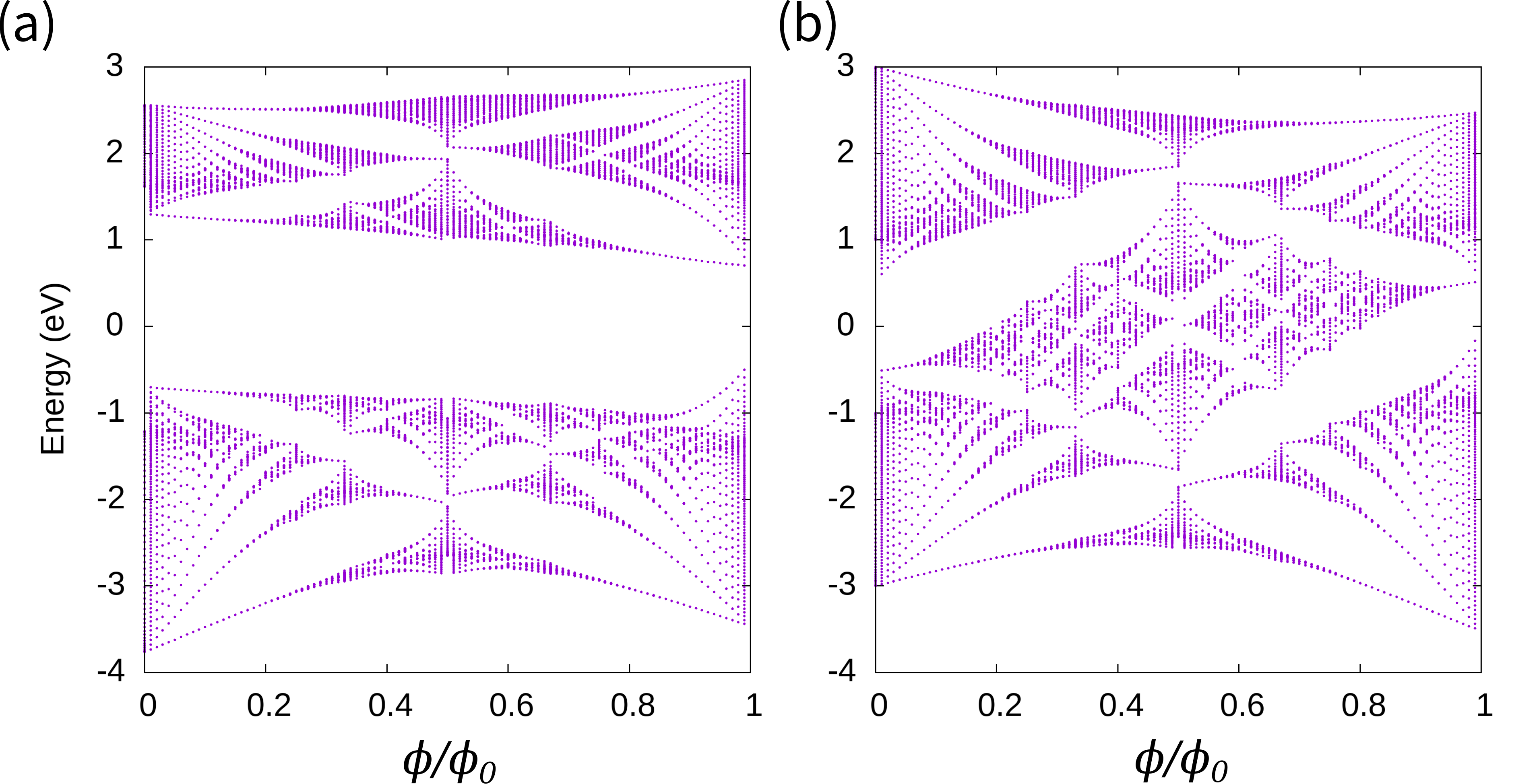}
	\caption{Hofstadter butterfly spectrum for a gapped two-band model with (a) trivial bands and (b) topological bands with the Chern numbers $C=+ 1$ in the lower band and $C=-1$ in the upper band. The Chern gap reaches the band edge in (b) as predicted by the Streda formula.}
	\label{fig:comphofs}
\end{figure}

We show the standard Hofstadter butterfly spectrum of a gapped phase with trivial bands in Fig.~\ref{fig:comphofs}(a), computed here for a two-band system with each band with finite bandwidth. In particular, the gap of the zero-field phase is preserved at a finite field. On the contrary, the Hofstadter spectrum of the nontrivial Chern phase is gapless, see Fig.~\ref{fig:comphofs}(b) for the two-band system now with $C=\pm 1$ Chern numbers. The closing of the gap is here explained by the change of the filling factor of the principal gap as a function of the magnetic flux, \ie according to the Streda formula $\nu = C \phi/2+1/2$ for a filling $\nu_0=1/2$ at zero fields \cite{Dana1985e}.   

Similarly, while the Hofstadter spectrum of the phase with trivial Euler topology ($E_I = E_{II} = 0$) is gapped, we show that the nontrivial Euler phases exhibit gapless Hofstadter spectra with the crossing of the Landau levels at half-filling ($\nu=1/2$), \ie within the gap of the zero-field phases, at a finite magnetic flux. In this work, we identify several qualitative features of the Hofstadter spectrum that relate to the finite Euler classes $E_{I,II}$ of the phases at zero fields. The Hofstadter spectrum, band structures and the Wilson loop calculations are performed with the open-source package WannierTools \cite{WU2017}.

\section{Flat-band limit}

In the limit of small flux,
the effect of an external magnetic field $B$ on the energy levels takes the semi-classical form \cite{PhysRevB.91.214405,PhysRevLett.99.197202,Alexandradinata_semiclassical,Alexandradinata_LQ} 
\begin{equation}
    \epsilon_{n,B}(k) \approx \epsilon_{n,0} (k) + m_n(k)\,B,
\label{eq_mag_response}
\end{equation}
where $\epsilon_{n,0}(k)$ is the energy eigenvalue at zero flux, and $m_n(k)$ describes the orbital magnetic susceptibility of the $n$-th band. In the case of TBG, it has been shown that $m_n(k)$ is related to the band topology at zero fields \cite{Wu2020}. More generally, the orbital magnetic susceptibility has contributions both from Berry curvature \cite{thonhauser2005orbital,resta2010electrical,THONHAUSER2011theory} and from the quantum geometry of the bands \cite{piechon2016geometric,hwang2011geometric}. By minimizing the effect of dispersion, the flatness of the bands thus makes the Landau levels a good probe of the topology and the quantum geometry of the bands \cite{rhim2020quantum}.

\begin{figure}
	\centering
	\includegraphics[width=0.9\linewidth]{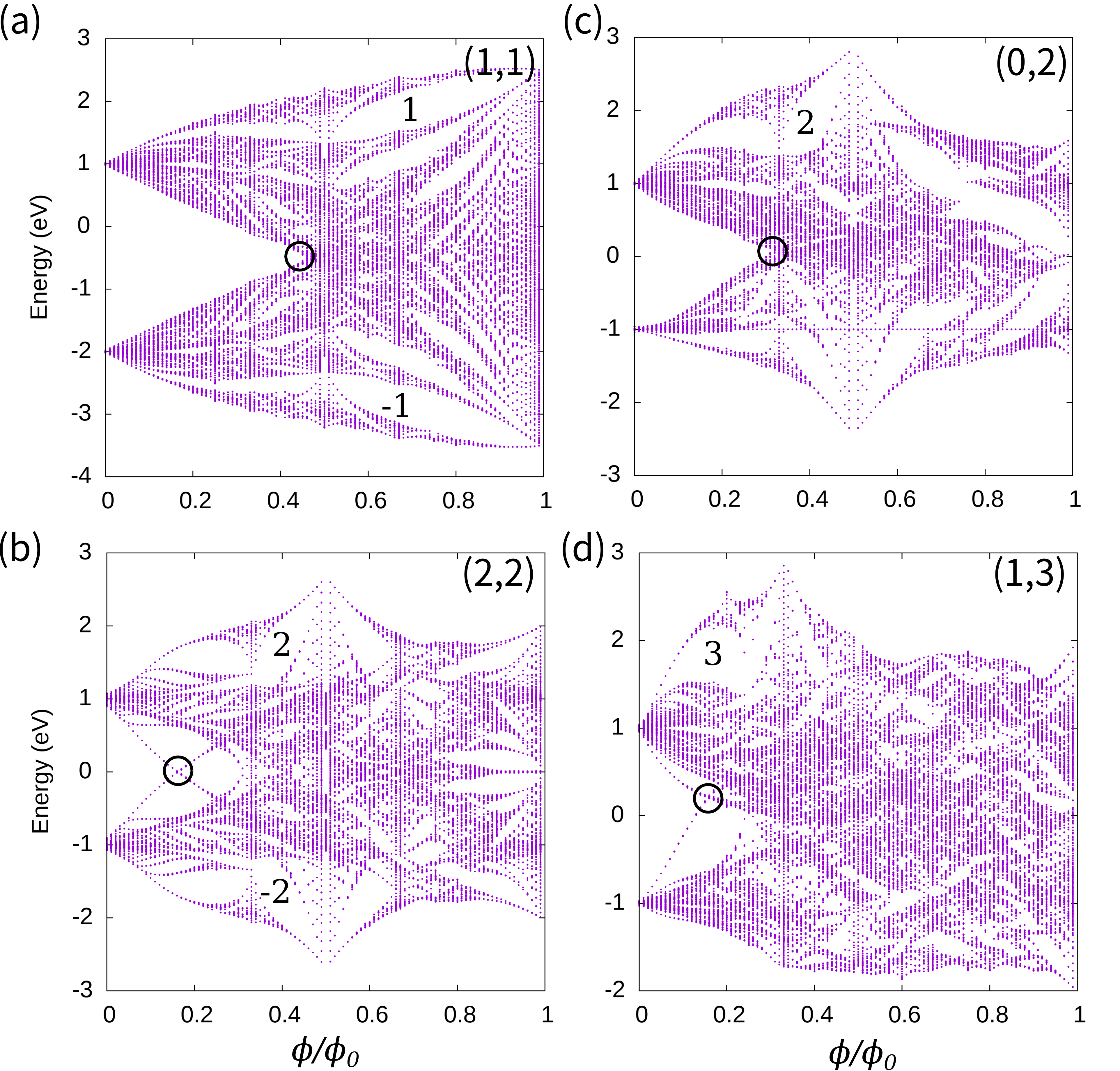}
	\caption{Hofstadter butterfly spectra calculated in the flat-band limit for the balanced phases with Euler classes ($E_I$,$E_{II}$) of (a) (1,1), (b) (2,2) and the imbalanced phases with ($E_I$,$E_{II}$) of (c) (0,2) and (d) (1,3). The Chern numbers of the main magnetic band gaps are written on the pictures.}
	\label{fig:HB_flat}
\end{figure}

Figure~\ref{fig:HB_flat} shows the Hofstadter butterfly spectrum for different Euler phases in the flat-band limit.
We find the qualitative trend that the crossing point of the Landau levels at half-filling moves toward zero for higher Euler classes, {\it i.e.}~the minimum magnetic field at which the Landau levels cross, $\phi_{\mathrm{cross}}$, decreases with increasing Euler classes.

We now give the rationale for the gap-closing of the Hofstadter butterfly. 
Starting with the balanced phases, Fig.~\ref{fig:HB_flat}(a,b), we show in Appendix C that every $C_{2z}\mathcal{T}$-symmetric Bloch Hamiltonian with {\it two-by-two} degenerate bands is {\it necessarily} symmetric under an effective basal mirror symmetry $m_z = C_{2z}I$, with $I$ the inversion symmetry and $C_{2z}$ a spinful $\pi$-rotation ({\it i.e.}~$C_{2z}^2=-1$) around the axis perpendicular to the basal plane of the two-dimensional system. The degenerated system is thus symmetric under the magnetic point group $2'/m = \{E,m_z,C_{2z}\mathcal{T},I\mathcal{T} \}$, with $[C_{2z}\mathcal{T}]^2=+1$ and $[I\mathcal{T}]^2=-1$, where the $I\mathcal{T}$ symmetry implies Kramers degenerate bands at all momenta. We conclude that the degenerate limit exists for all balanced Euler insulating phases without the need for fine-tuning. In other words, the degeneracy of the bands is always associated with a symmetry of the Hamiltonian such that it is not accidental (see Appendix C for a detailed exposition). We also find that the Chern basis Eq.~(\ref{eq_chern_basis}), {\it i.e.}~an eigenbasis of the balanced Hamiltonian in the flat-band limit, is an eigenbasis of the $m_z$ symmetry operator (Appendix C). There is thus a one-to-one correspondence between the Euler class and the mirror Chern number of the occupied bands, that is (Appendix C)
\begin{equation}
\label{eq_euler_chern}
    E_I = E_{II} = -C^{(-\mathrm{i})} = C^{(\mathrm{i})} \in \mathbb{Z} . 
\end{equation}
Interestingly, the enrichment of $C_{2z}\mathcal{T}$-symmetric phases with $m_z$ symmetry implies that each two-band vector subbundle becomes oriented with the signed Euler class as topological invariant, {\it i.e.}~$E^{(-\mathrm{i})}_I \in \mathbb{Z}$, by virtue of attaching a fixed orientation to each mirror-eigenvalue sector (Appendix C). Given that $m_z$-symmetry is preserved at finite magnetic field $B_z$, the crossings between Landau levels of distinct mirror eigenvalues are protected by symmetry. We now can derive the qualitative trend as a function of the Euler class from the Streda formula 
$\nu = C^{(\mp\mathrm{i})} \phi/2 + \nu_0$ with $\nu_0 = 1/2$ the filling at zero flux \cite{Dana1985e}. 
Indeed, the subband with Chern number $C^{(\mp\mathrm{i})}$ must reach the band edge ($\nu=0,1$) at $\phi=1/|C^{(\mp\mathrm{i})}|$, which gives an upper bound for the gap-closing flux in each $m_z$-eigensector (see Fig.~\ref{fig:comphofs}(b) showing the Hofstadter spectrum of a generic Chern insulating phase with $C=+1$ in the lower band). More precisely, the occupied Landau levels at half-filling with a Chern number $C_v = \max\{ C^{(-\mathrm{i})}_I, C^{(\mathrm{i})}_I \}$ at zero flux must reach the filling $\nu=C_v/2+1/2$ at the flux $\phi=1$ ({\it e.g.}~$\nu=1$ if $C_v = 1$), while the conduction Landau levels at half-filling with a Chern number $C_c = \max\{ C^{(-\mathrm{i})}_{II}, C^{(\mathrm{i})}_{II} \}$ at zero flux must reach the filling $\nu = -C_c/2+1/2$ (note the sign change for the conduction bands) at the flux $\phi=1$ ({\it e.g.}~$\nu=0$ if $C_c = 1$). Furthermore, if $C_v = C^{(-\mathrm{i})}_I > 0$ (and thus $C_v = C^{(\mathrm{i})}_I > 0$), then $C_c = C^{(\mathrm{i})}_{II} > 0$ (and $C_c = C^{(-\mathrm{i})}_{II} > 0$), since $C^{(\mp\mathrm{i})}_{I} + C^{(\mp\mathrm{i})}_{II} = 0$.
We hence conclude that Landau levels of distinct $m_z$-eigenvalues must cross at half-filling (see also \cite{Herzog-Arbeitman2020a}), with the trend through Eq.~(\ref{eq_euler_chern}) of a smaller gap-closing flux for a higher Euler class. This is in agreement with the numerical results.

We now consider the imbalanced phases ($\vert E_I\vert  \neq \vert E_{II}\vert $) in the flat-band limit, shown in Fig.~\ref{fig:HB_flat}(c,d), where the same trend is observed. Contrary to the balanced case, there is no effective mirror symmetry. This implies that the two-by-two band degeneracy requires fine-tuning. We find that the exact degeneracy, similarly to the flat-band limit, requires infinite-range hopping terms in the tight-binding model. In practice, it can be achieved in very good numerical approximation by keeping hopping terms up to sufficiently far neighbours, see Appendix A. The absence of effective mirror symmetry in the imbalanced Euler phases leaves unexplained the stability of the Landau level crossing at half-filling [Fig.~\ref{fig:HB_flat}(c,d)]. 
Nevertheless, by making use of the Chern basis Eq.~(\ref{eq_chern_basis}) as the eigenbasis (permitted by the two-by-two degeneracy of the bands), we can still decompose the bands at zero magnetic fields into two decoupled imbalanced Chern insulators, {\it i.e.}~$H = H^{+} \oplus H^{-}$ with $C^{\pm}_I = \pm E_I$ and $C^{\pm}_{II} = \mp E_{II}$. The stability of the gaplessness of the Hofstadter spectrum, Fig.~\ref{fig:HB_flat}(c,d), suggests that the effect of the magnetic field introduced via the Peierls substitution preserves the decoupling between the two Chern sectors, even though there is no global symmetry of the Hamiltonian protecting the decoupling (like $m_z$ in the balanced case). 

\section{Dispersive bands}\label{sec_dis_band}

\begin{figure}
	\includegraphics[width=8.6cm]{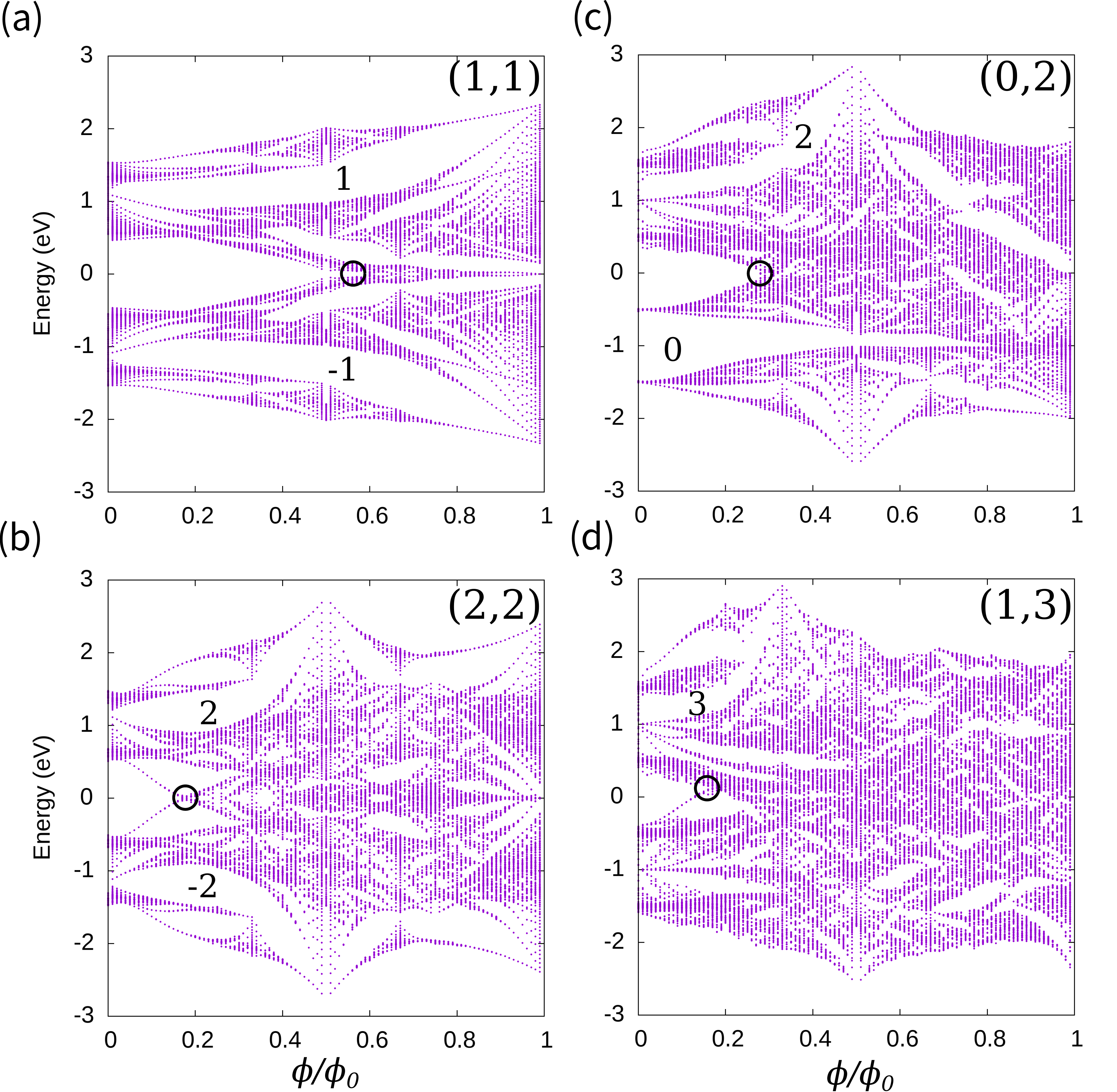}
	\caption{Hofstadter butterfly spectra calculated for the Euler insulating phases away from the flat-band and degenerate limit
	for Euler classes (a) (1,1), (b) (2,2), (c) (0,2) and (d) (1,3).
 While the effective $m_z$-symmetry is broken, the balanced phases all satisfy the hidden symmetry (see text). The magnetic subgap Chern numbers are bounded from below by the Euler classes.}
	\label{fig:HB}
\end{figure}

We are now ready to address the more general situation of dispersive (non-flat) and non-degenerate Euler insulating phases [\eg Fig.\;\ref{fig:bandwcc}(a)]. 
Any adiabatic perturbation of Eq.~(\ref{eq_H_imb}) removes the degeneracy and the flatness of the bands while preserving the Euler class topology. From the Hofstadter spectra shown in Fig.~\ref{fig:HB}, we readily find that the gaplessness at half-filling remains a feature of the nontrivial Euler insulating phases, both for the balanced Fig.~\ref{fig:HB}(a,b) and imbalanced Fig.~\ref{fig:HB}(c,d) phases. 
This is somehow surprising since the non-degenerate balanced phases do not preserve the effective $m_z$-symmetry ({\it i.e.}~there is no mirror Chern number), and the Chern basis [Eq.~(\ref{eq_chern_basis})] are not eigenvectors of the Hamiltonian anymore. We hence would conclude that, in principle, the crossing of the Landau level branches in the gap at half-filling is not protected, as was reported in Ref.~\cite{Herzog-Arbeitman2020a} for the $E_I = E_{II}=1$ case. We give below an explanation for Landau-level crossings in the phases with nonzero Euler classes at zero-field.

We first resolve the apparent contradiction, \ie the observed Landau-level crossings without symmetry protection, in the case of the balanced phases. By systematically probing all perturbations of the Euler insulating phases allowed by $C_{2z}\mathcal{T}$ symmetry, we find that only the term $h_{yy}$ added to $\widetilde{H}_{\mathrm{bal}}[\boldsymbol{n}] $ [Eq.\;(\ref{eq_bal_H})] controls the gapping of the Hofstadter spectrum, see Fig.~\ref{fig:HB_gapped}(a,b) obtained with a constant term $ h_{yy} = \delta > 0$ added adiabatically. (See Appendix D, for a detailed discussion of all the symmetry-allowed perturbations.) Since there is no global symmetry that can account for the vanishing or non-vanishing of this term, we call the condition $h_{yy} = 0$ a {\it hidden symmetry} of the balanced Euler insulating phases described by Eq.\;(\ref{eq_bal_H}). 
We note that under a change of orbital basis of the Bloch Hamiltonian, the term that controls the hidden symmetry must be changed accordingly. We emphasize that all the balanced Euler phases at zero flux shown in Fig.~\ref{fig:HB}(a,b) (and in \cite{Supple}) satisfy the hidden symmetry, {\it i.e.}~all the terms $h_{ij}$ in Eq.\;(\ref{eq_ten_terms}) are non-zero except $h_{yy}$.

The imbalanced phases on the contrary are mainly unaffected by $C_{2z}\mathcal{T}$-preserving perturbations, exhibiting a robust gapless Hofstadter spectrum, as one can see in the results Fig.~\ref{fig:HB_gapped}(d-f) obtained for $h_{yy} \neq 0$. 
This can be understood by noting that the flat degenerate imbalanced Bloch Hamiltonian, Eq.~(\ref{eq_H_imb}) with $q,q'\neq 0$, already has all the nine independent terms $h_{ij}$ in Eq.\;(\ref{eq_ten_terms}) nonzero and without relations between them, such that a further perturbation in $h_{yy}$ does not lead to a qualitative change of the spectrum.

\begin{figure}
	\centering
	\includegraphics[width=8.6cm]{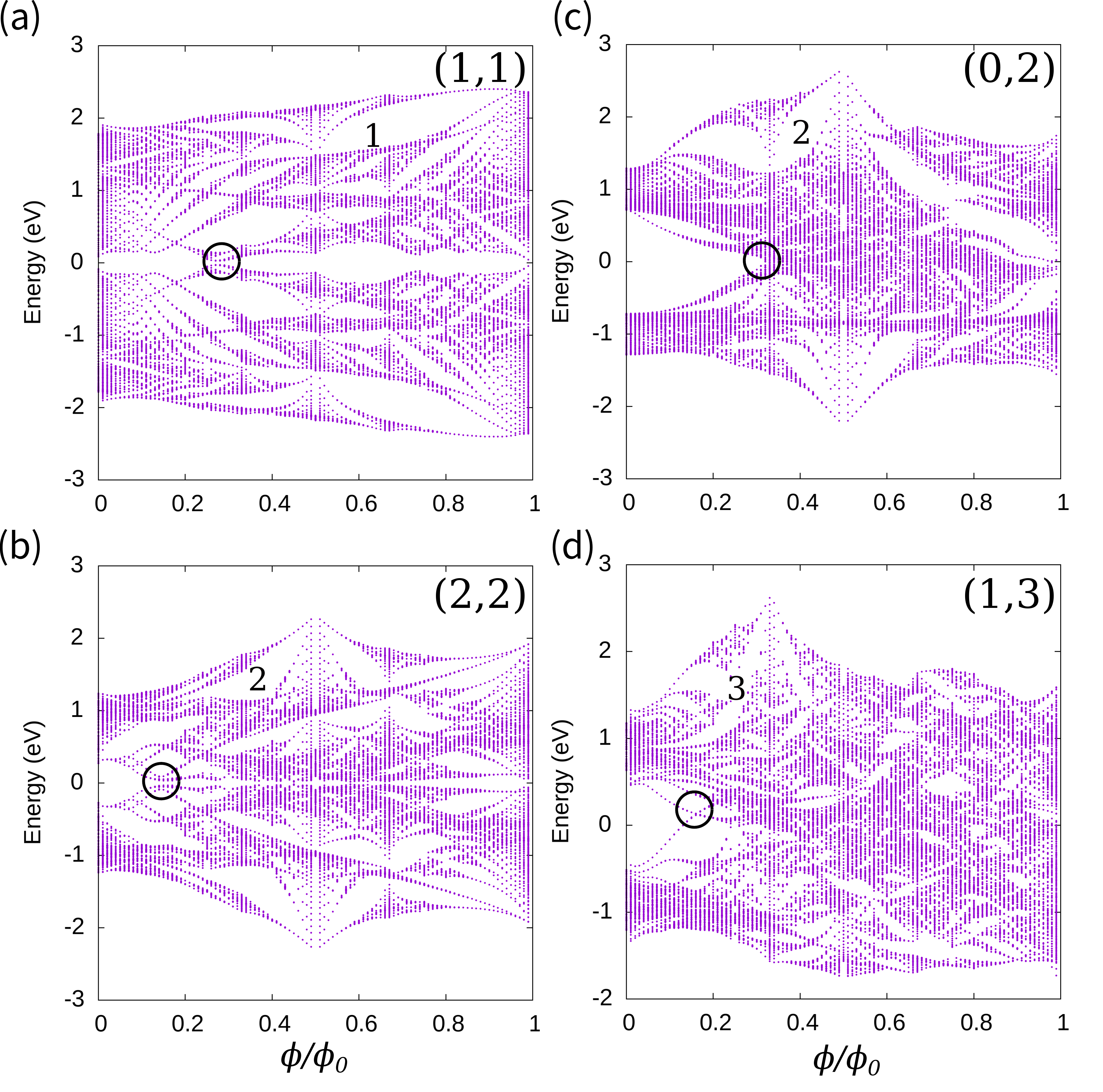}
	\caption{Effect of breaking the hidden symmetry on the Hofstadter butterfly spectra for balanced (a) (1,1), (b) (2,2) and imbalanced (c) (0,2), (d) (1,3) phases.
The imbalanced phases remain gapless. The magnetic subgap Chern numbers are still bounded from below by the Euler classes.
	}
	\label{fig:HB_gapped}
\end{figure}

\subsection{Magnetic sub-gaps}

When the nodal points, located at quarter fillings $\nu = 1/4,3/4 $, are well separated in energy from the rest of the bands, we can easily identify the Landau levels originating from the nodes at a small magnetic flux. We find that the number of stable nodal points contained in each two-band subspace gives a lower bound for the Chern numbers of the magnetic sub-gaps above and below these Landau levels, {\it i.e.}~$2 \vert E_{\alpha}\vert \leq \vert C_{\nu=(2 [\alpha]-1)/4} \vert$ with $[\alpha]$ defined by $[I]=1$ and $[II]=2$. 
Indeed, the Chern numbers of the magnetic sub-gaps can be increased by adding unstable nodes, such as e.g.~in graphene, while the Euler class dictates the minimal number of nodes to be $2 \vert E_{\alpha}\vert$ in each two-band subspace.

\section{Discussion}

We briefly discuss the difference between the Landau levels of Euler insulators with the Landau levels of mirror Chern insulators and time-reversal symmetric topological insulators (\ie the Kane-Mele $\mathbb{Z}_2$ quantum spin Hall phases).
First of all, despite the existence of a mirror ($m_z$) Chern number in the (flat) two-by-two degenerate Euler phases, the Euler insulators, in general, are different from mirror Chern insulators by their fragility. That is, while the mirror Chern topology is stable, the $\mathbb{Z}$ Euler topology of two-band subspaces is reduced to the $ \mathbb{Z}_2$ second Stiefel-Whitney class topology under the addition of trivial bands \cite{po2018fragile,Bouhon2018,BJY_nielsen,Lian2020b}. 
Furthermore, the $\mathbb{Z}_2$ Stiefel-Whitney insulators, which have no topological edge states, are also distinct from the Kane-Mele $\mathbb{Z}_2$ phases, with topological helical edge states. In the context of the Hofstadter spectrum, time-reversal symmetry is broken by the magnetic field and, if no other symmetry is present, the $\mathbb{Z}_2$ Kane-Mele phases exhibit a gapped Hofsdtater spectrum \cite{Herzog-Arbeitman2020a}.

We now discuss the potential candidates for observing the manifestations of Euler topology in the Hofstadter spectrum. Since the Landau levels rely on the effect of the magnetic field, our findings can be more naturally realized in electronic systems than in charge-neutral systems, such as optical lattices \cite{UnalQuench} or acoustic metamaterials \cite{jiang2021observation}. 
In that sense, the family of materials with moir{\'e} superlattices would be of interest. 
Indeed, moir{\'e} superlattices in twisted heterostructures provide the possibility to realize and tune the  fragile topology, while the large supercell facilitates the measurement of the Hofstadter butterfly that requires a very high magnetic flux per unit cell. We for instance propose the $M+N$ twisted multilayer graphene built by stacking the $M$-layer and $N$-layer graphene multilayers with a twist as a platform for realizing Euler insulators with arbitrary topological charge, since in such superlattices the flat bands can carry $(M-N)$ Chern numbers \cite{Liu2019a,zhang2020chiral}. 

To conclude, we have studied the response of $C_2\mathcal{T}$ symmetric fragile topological insulators to external magnetic fields with tight-binding models hosting a variety of balanced and imbalanced Euler insulating phases. We have shown that the Hofstadter energy spectrum is affected qualitatively by the topological Euler class, especially in the flat-band limit. Our results also provide an insight for the study of topological flat-band systems with non-trivial quantum metrics, such as the topologically bounded superfluid weight found in twisted multilayered systems \cite{Xie2020_superfluid} or the divergence found in the Landau levels of anomalous flat bands \cite{rhim2020quantum}, generalizing to the very rich, yet mainly unexplored, landscape of higher Euler class phases.

\medskip{}

\emph{Acknowledgement-}Y.G. acknowledges support by the Swiss NSF (grant
No. 172543) Computations were performed at the Swiss National Supercomputing
Centre (CSCS) under projects Nos. s832 and s1008 and the facilities
of Scientific IT and Application Support Center of EPFL.

Y.G. and A. B. initiated the project and Y.G. obtained the numerical results. A. B. obtained the analytical results and wrote the manuscript. All authors discussed the results.  

\appendix

\section{The modelling of Euler insulating phases}\label{sec:modeling_ET}

\subsection{The geometric approach of Refs.~\cite{BouhonGEO2020,abouhon_EulerClassTightBinding}}

We here briefly review the construction of the homotopy representative Hamiltonian for the four-band Euler insulating phases at half-filling following Ref.~\cite{BouhonGEO2020,abouhon_EulerClassTightBinding}. 

The spectral decomposition of the $4\times 4$ real symmetric Hamiltonian $\widetilde{H}$, {\it i.e.}~$\widetilde{H}  u_n =\epsilon_n u_n $ with the eigenvalue $\epsilon_n$ and the eigenvector $u_n \in \mathbb{R}^{4}$ for $n=1,\dots,4$, gives $\widetilde{H} = R\cdot  D \cdot R^T$ with $R = (u_1~u_2~u_3~u_4) \in \mathsf{SO}(4)$ the matrix of real eigenvectors and $D = \mathrm{diag}(\epsilon_1,\epsilon_2,\epsilon_3,\epsilon_4)$ the matrix of energy eigenvalues. In the following we set $\epsilon_1 = \epsilon_2 = -\epsilon$ and $\epsilon_3 = \epsilon_4 = \epsilon > 0$. 

From the spectral decomposition and the degeneracy of the energy levels, we readily have that $\widetilde{H}$ is invariant under any gauge transformation $R\rightarrow R G$ with $G=G_v \oplus G_c$ and $G_v,G_c \in \mathsf{O}(2)$, such that $\det G = \det (G_v) \det (G_c) = 1$. Defining the corresponding left coset $[R] = \{R G \vert G \in \mathsf{S}[\mathsf{O}(2)\times \mathsf{O}_2]\}$, we thus find that the Hamiltonian is an element of the real {\it unoriented} Grassmannian as $[R] \in \mathsf{SO}(4)/\mathsf{S}[\mathsf{O}(2)\times \mathsf{O}(2)] = \mathrm{Gr}^{\mathbb{R}}_{2,4}$. 

We are here excluding non-orientable phases characterized by $\pi$-Berry phases along the two non-contractible loops of the Brillouin zone. While the Hamiltonian defines an {\it orientable} vector bundle (see Appendix \ref{ap_top_Euler} below) \cite{BouhonGEO2020}, it is convenient to first seek an element of the real \textit{oriented} Grassmannian $\widetilde{\mathrm{Gr}}^{\mathbb{R}}_{2,4} = \mathsf{SO}(4)/[\mathsf{SO}(2)\times \mathsf{SO}(2)]$ to construct the Hamiltonian. This allows us to take advantage of the diffeomorphism $\widetilde{\mathrm{Gr}}^{\mathbb{R}}_{2,4} \cong \mathbb{S}^2 \times  \mathbb{S}^2$. Starting from the explicit parametrization of $R$ as a generic element of $\mathsf{SO}(4)$, the reduction to the oriented Grassmannian is then carried out through the Pl{\"u}cker embedding permitting the representation of the Grassmannian as a $4$-dimensional manifold subspace of a $6$-dimensional vector space (the second exterior power of $\mathbb{R}^4$), {\it i.e.}
\begin{equation}
\label{eq_Pembedding}
    \iota : \widetilde{\mathrm{Gr}}^{\mathbb{R}}_{2,4} \xhookrightarrow{}  \bigwedge\nolimits^2(\mathbb{R}^{4}) : [R] \mapsto (\boldsymbol{n}_+,\boldsymbol{n}_-) \in \mathbb{S}^2_+ \times \mathbb{S}^2_- ,
\end{equation}
where 
\begin{equation}
 \boldsymbol{n}_{\pm}(\phi_{\pm},\theta_{\pm}) = (\cos \phi_{\pm} \sin\theta_{\pm},\sin\phi_{\pm} \sin\theta_{\pm},\cos\theta_{\pm}),
\end{equation} 
are the unit vectors on the two unit spheres $\mathbb{S}^2_{\pm}$ living in two perpendicular 3-dimensional vector subspaces of $\bigwedge\nolimits^2(\mathbb{R}^{4})$. Since the second arrow in Eq.~(\ref{eq_Pembedding}) is a bijection, we write the representative of each coset $[R]$ as $R(\boldsymbol{n}_+,\boldsymbol{n}_-)$, and the Euler Hamiltonian is readily given by 
\begin{multline}
    \widetilde{H}_E[\boldsymbol{n}_+,\boldsymbol{n}_-] =\\ R(\boldsymbol{n}_+,\boldsymbol{n}_-)\cdot \left(\begin{array}{cc}
        -\epsilon\, \mathbb{1} & \mathbb{0} \\
        \mathbb{0} & \epsilon\, \mathbb{1} 
    \end{array}\right)\cdot R(\boldsymbol{n}_+,\boldsymbol{n}_-)^T.
\end{multline}
See the Mathematica code of Ref.~\cite{abouhon_EulerClassTightBinding} for the explicit expression of $R(\boldsymbol{n}_+,\boldsymbol{n}_-)$ as a function of the four spherical angles $(\phi_{+},\theta_{+},\phi_{-},\theta_{-})$. Defining 
\begin{equation}
    \begin{aligned}
        \boldsymbol{n}&=\boldsymbol{n}_+(\phi_+,\theta_+) ,\\
        \boldsymbol{n}' &= \boldsymbol{n}_-(\phi_-+\pi/2,\theta_-+\pi/2),
    \end{aligned}
\end{equation}
and writing the components $\boldsymbol{n}^{(\prime)} = (n_1^{(\prime)},n_2^{(\prime)},n_3^{(\prime)})$, the Euler Hamiltonian is then \cite{BouhonGEO2020,abouhon_EulerClassTightBinding}
\begin{equation}
\label{eq_H_general}
\begin{aligned}
    \widetilde{H}_E[\boldsymbol{n},\boldsymbol{n}'] &= n_1' (
        -n_1 \Gamma_{zz} + n_2 \Gamma_{zx} + n_3 \Gamma_{x0}
    )& \\
     &- n_2' (
        +n_1 \Gamma_{xz} - n_2 \Gamma_{xx} + n_3 \Gamma_{z0}
    )& \\
    &+ n_3' (
       + n_1 \Gamma_{0x} + n_2 \Gamma_{0z} - n_3 \Gamma_{yy}
    )&,
\end{aligned}
\end{equation}
with $\Gamma_{ij} = \sigma_i\otimes \sigma_j$ and the Pauli matrices $\{\sigma_i\}_{i=x,y,z}$, and with $\sigma_0 = \mathbb{1}$.

\subsubsection{Homotopy classification}

Considering the unit vector $\boldsymbol{n}$ as a mapping from a base sphere $\mathbb{S}^2_0$ to a target sphere $\mathbb{S}^2$, {\it i.e.}~$\boldsymbol{n} = \boldsymbol{n}(\phi(\phi_0,\theta_0),\theta(\phi_0,\theta_0))$, we define the skyrmion number
\begin{equation}
\label{eq_Skyrmion}
    W[\boldsymbol{n}] =  \displaystyle \dfrac{1}{4\pi} \int_{\mathbb{S}^2_0} d\phi_0 d\theta_0 ~ \boldsymbol{n} \cdot (\partial_{\phi_0}\boldsymbol{n}   \times \partial_{\theta_0}\boldsymbol{n}) \in \mathbb{Z},
\end{equation}
that counts the number of times $\boldsymbol{n}$ wraps the target sphere $\mathbb{S}^2$ as we cover the base sphere $\mathbb{S}^2_0$ one time. By setting
\begin{equation}
\begin{aligned}
    \boldsymbol{n}_q &= (\cos(q\phi_0)\sin\theta_0 , \sin(q\phi_0)\sin\theta_0,\cos\theta_0),\\
    \boldsymbol{n}'_{q'} &= (\cos(q'\phi_0)\sin\theta_0 , \sin(q'\phi_0)\sin\theta_0,\cos\theta_0) ,
\end{aligned}
\end{equation}
we readily obtain
\begin{equation}
    W[\boldsymbol{n}_q] = q,~ \mathrm{and} ~
    W[\boldsymbol{n}'_{q'}] = q',
\end{equation}
in terms of which the homotopy classification of $\widetilde{H}[\boldsymbol{n}_q,\boldsymbol{n}'_{q'}]$ in Eq.~(\ref{eq_H_general}) is defined, since
\begin{equation}
\begin{aligned}
   \pi_2[\widetilde{\mathrm{Gr}}^{\mathbb{R}}_{2,4}] &= \pi_2[\mathbb{S}^2\times \mathbb{S}^2] 
   = \pi_2[\mathbb{S}^2] \oplus \pi_2[\mathbb{S}^2]\\
   &= \mathbb{Z}\oplus \mathbb{Z} \ni (q,q').
\end{aligned}
\end{equation}

\subsection{Topology of the Bloch Hamiltonian}\label{ap_top_Euler}

So far, we have not specified the parameter base space of the Hamiltonian. Considering a two-dimensional crystalline system, we aim at a Bloch Hamiltonian $\widetilde{H}(\boldsymbol{k})$ parametrized by a momentum vector $\boldsymbol{k}=(k_1,k_2)$ inside the Brillouin zone $[-\pi,\pi)^2 \cong \mathbb{T}^2$.  

Preceding the previous construction by a projection of the Brillouin zone onto the base sphere $\mathbb{S}_0$, {\it i.e.} 
\begin{equation}
    \mathbb{T}^2 \rightarrow \mathbb{S}^2_0 \rightarrow \mathbb{S}^2 \times \mathbb{S}^{2\prime} : \boldsymbol{k} \mapsto \boldsymbol{n}_0(\boldsymbol{k}) \mapsto (\boldsymbol{n}_q(\boldsymbol{k}),\boldsymbol{n}'_{q'}(\boldsymbol{k})),
\end{equation}
we obtain an explicit parametrization of the Hamiltonian Eq.~(\ref{eq_H_general}) as a Bloch Hamiltonian for all the homotopy classes, {\it i.e.}
\begin{equation}
    \widetilde{H}_E[\boldsymbol{n},\boldsymbol{n}']\rightarrow \widetilde{H}_E[\boldsymbol{n}_q(\boldsymbol{k}),\boldsymbol{n}'_{q'}(\boldsymbol{k})] .
\end{equation}

Writing the Euler classes (see main text) of the occupied and unoccupied bands $E_I$ and $E_{II}$, respectively, we obtain the homotopy classification of the two-dimensional \textit{orientable} (excluding $\pi$-Berry phases) four-band Euler insulating phases through 
\begin{equation}
\label{eq_winding}
        E_I =  q-q'  ,~
        E_{II} =  q+q' .
\end{equation}
Importantly, the homotopy classification $\pi_2[\mathsf{Gr}_{2,4}^{\mathbb{R}}]= \pi_2[\widetilde{\mathsf{Gr}}_{2,4}^{\mathbb{R}}] = \pi_2[\mathbb{S}^2\times \mathbb{S}^2] = \mathbb{Z} \oplus \mathbb{Z}$, assumes the constraint of a fixed base point (by definition of the homotopy groups). However, Bloch Hamiltonians do not fix a base point, which allows the nontrivial action of the generator of the first homotopy group on the second homotopy group \cite{BouhonGEO2020}. Writing $\pi_1[\mathsf{Gr}_{2,4}^{\mathbb{R}}] = \mathbb{Z}_2 = \{[x],[\ell]\}$, where $[x]$ is the class of loops that can be shrunk to a point and the generator $[\ell]$ is the class of loops that cannot be shrunk to a point, the action of $[\ell]$ on the second homotopy group is represented by the deformation of a reference point of $\mathsf{Gr}_{2,4}^{\mathbb{R}}$ over a nontrivial loop in $\mathsf{Gr}_{2,4}^{\mathbb{R}}$. This induces the flip of both Euler classes, \ie 
\begin{equation}
    (E_I,E_{II}) \xrightarrow[]{[\ell]}(-E_I,-E_{II}),
\end{equation}
leading to a reduction of the classification \cite{BouhonGEO2020}
\begin{multline}
    (q,q') \in \mathbb{Z}^2 \longrightarrow  \\
     (E_I,E_{II}) \in \left\{(a,b) \in \mathbb{Z}^2 \vert (a,b)\sim(-a,-b) \right\},
\end{multline}
(this captures the distinction between the topology of \textit{oriented} and \textit{orientable} spaces). Moreover, it can be shown that the homotopy classification of the balanced phases ($\vert E_I\vert=\vert E_{II}\vert$) is further reduced due to the existence of an adiabatic transformation between the phases $(E_I,E_{II})=(q,q)$ and $(-q,q)$ \cite{bouhon2022multigap}.

We finally note the sum rule 
\begin{equation}
    E_{I}+E_{II} = 0\,\mathrm{mod}\,2,  
\end{equation}
which guarantees the cancellation of the second Stiefel-Whitney class over all the bands, {\it i.e.}~$w_{2,I} + w_{2,II} = 0 \,\mathrm{mod}\,2$.

\subsubsection{Balanced and imbalanced phases}

The above homotopy classification allows us to distinguish two types of phases, the balanced phases with $\vert E_I \vert = \vert E_{II}\vert $, and the imbalanced phases with $\vert E_I\vert \neq \vert E_{II}\vert$. The balanced phases are characterized by having one zero skyrmion number, {\it i.e.} either $q=0$ or $q'=0$, while the imbalanced phases are characterized by having two nonzero skyrmion numbers, {\it i.e.} $\vert q \vert , \vert q'\vert>0$.

In Appendix \ref{sec:com_real} and \ref{sec:mz}, we rederive in detail the general form and the topology of the Bloch Hamiltonian for the balanced phases with degenerate bands by starting from a system with spinful basal mirror symmetry. Indeed, we prove in the Appendix \ref{ap_thm_deg_balanced_mirror} that a balanced Euler insulating phase has degenerate bands if and only if it has spinful basal mirror symmetry.

\subsection{Tight-binding Hamiltonian}

Once the homotopy representative Hamiltonian has been parametrized in terms of the points of the Brillouin zone, as in $\widetilde{H}_E[\boldsymbol{n}_q(\boldsymbol{k}),\boldsymbol{n}'_{q'}(\boldsymbol{k})]$, we get a tight-binding Bloch Hamiltonian by expanding each term as a Fourier series, {\it i.e.}
\begin{equation}
\begin{aligned}
\label{eq_Fourier}
    \left(\widetilde{H}_{E}[\boldsymbol{n}_q(\boldsymbol{k}),\boldsymbol{n}'_{q'}(\boldsymbol{k})]\right)_{ab} &=  \sum\limits_{l_x,l_y\in \mathbb{Z}} 
    \left[a^{(ab)}_{(l_x,l_y)} \cos( l_x k_x + l_y k_y)  \right.\\ 
    &\left.
    + b^{(ab)}_{(l_x,l_y)} \sin( l_x k_x + l_y k_y) \right].
\end{aligned}
\end{equation}
In practise, we only need to keep a finite number $K$ of terms, such that $-K \leq l_x,l_y \leq K$,  since the hopping parameters $\left\{a^{(ab)}_{(l_x,l_y)},b^{(ab)}_{(l_x,l_y)}\right\}$ decrease rapidly with the distance $\sqrt{l_x^2+l_y^2}$.

The explicit tight-binding models used in this work have been retrieved from Ref.~\cite{abouhon_EulerClassTightBinding} which provides a Mathematica notebook that generates four-band (and three-band) tight-binding models for arbitrary fixed Euler classes.

\subsubsection{Flat-band limit}

\begin{figure}
    \centering
    \includegraphics[angle=270,width=8.6cm]{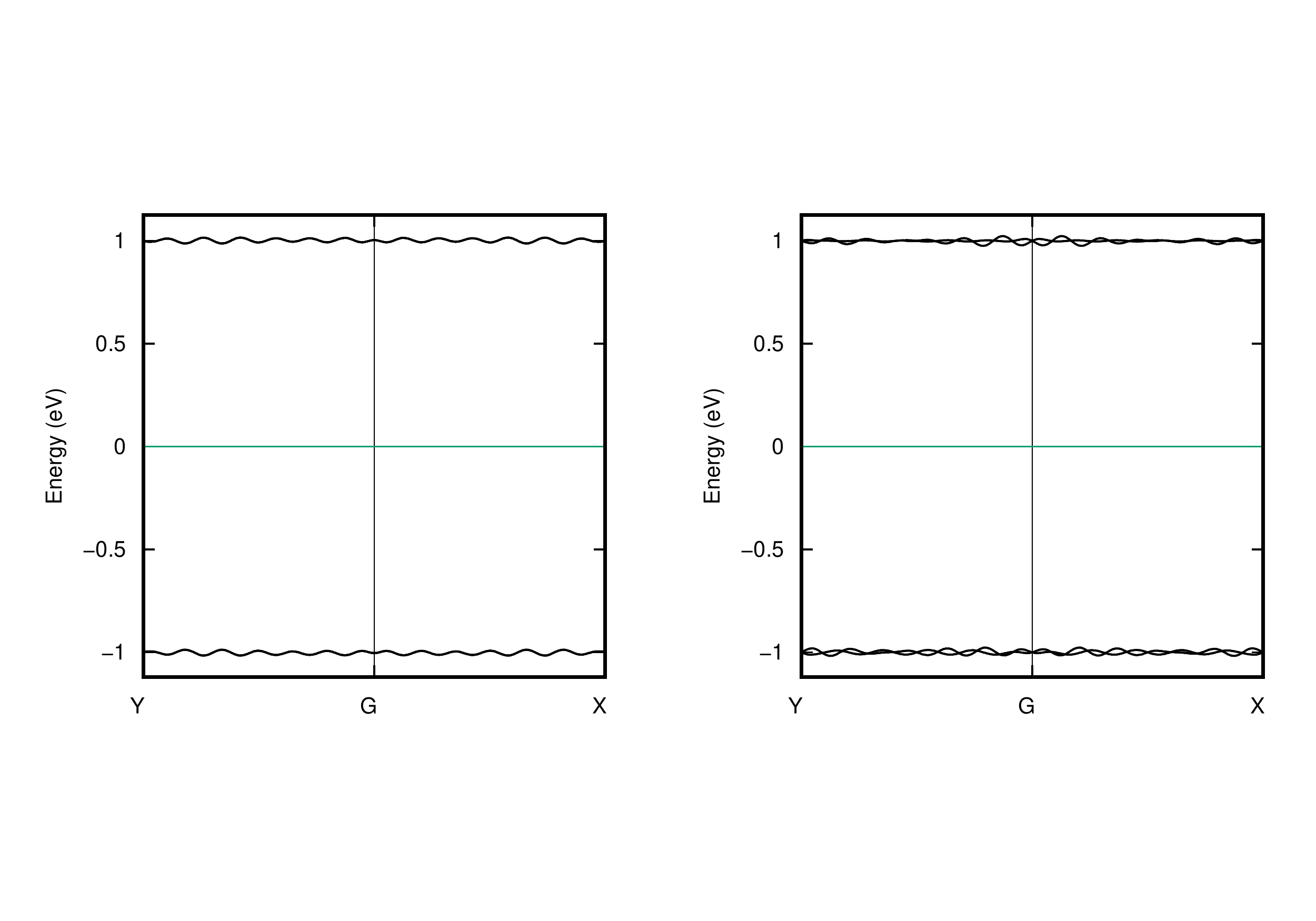}
    \caption{Band structures in the approximate flat-band limit for (a) the balanced phase $(E_I,E_{II}) = (2,2)$, and (b) the imbalanced phase $(E_{I},E_{I})=(1,3)$, obtained for $K=12$ in Eq.~(\ref{eq_Fourier}).}
    \label{fig:BS_flat_band}
\end{figure}

In our context, perfect flat bands would require to keep all hopping terms up to infinitely distant neighbors ({\it i.e.} $ -\infty \leq l_{x,y} \leq \infty$ in Eq.~(\ref{eq_Fourier})). However, we obtain a very good numerical approximation of the flat bands by keeping hopping terms up to $K\approx 12$, see {\it e.g.} Fig.~\ref{fig:BS_flat_band}(a) and (b).

\section{From complex to real basis}\label{sec:com_real}

Usually, the tight-binding models of physical systems with $C_{2}\mathcal{T}$-symmetry are not given in their real form. In order to fix ideas, let us start from the following Bloch-L{\"o}wdin orbital basis, composed of two $s$-orbitals located at the center of the unit cell, each taken with the two spin-1/2 components,   
\begin{multline}
    \vert \boldsymbol{\varphi} , \boldsymbol{k} \rangle
    = \left(\vert \varphi_{1,\uparrow}, \boldsymbol{k} \rangle ~
    \vert \varphi_{2,\uparrow}, \boldsymbol{k} \rangle ~
    \vert \varphi_{1,\downarrow}, \boldsymbol{k} \rangle ~
    \vert \varphi_{2,\downarrow}, \boldsymbol{k} \rangle
    \right),\\
    = \sum\limits_{\boldsymbol{R}} \mathrm{e}^{\mathrm{i} \boldsymbol{k}\cdot \boldsymbol{R}} 
    \left(\vert w_{1,\uparrow}, \boldsymbol{R} \rangle ~
    \vert w_{2,\uparrow}, \boldsymbol{R} \rangle ~
    \vert w_{1,\downarrow}, \boldsymbol{R} \rangle ~
    \vert w_{2,\downarrow}, \boldsymbol{R} \rangle
    \right) ,
\end{multline}
where $\boldsymbol{R}$ runs over all the Bravais vectors of the lattice, and $\langle \boldsymbol{r} \vert w_{\alpha} , \boldsymbol{R} \rangle = w_{\alpha} (\boldsymbol{r}-\boldsymbol{R})$ is the Wannier function of the orbital $\alpha$ localized at the lattice site $\boldsymbol{R}$. The Bloch Hamiltonian then reads, 
\begin{equation}
    \mathcal{H} = 
    \sum\limits_{\boldsymbol{k}} \vert \boldsymbol{\varphi} , \boldsymbol{k} \rangle H(\boldsymbol{k}) \langle \boldsymbol{\varphi} , \boldsymbol{k} \vert ,
\end{equation}
with 
\begin{equation}
\begin{aligned}
    H(\boldsymbol{k}) &= \left( 
        \begin{array}{cc}
            H_{\uparrow\uparrow}(\boldsymbol{k}) & H_{\uparrow\downarrow}(\boldsymbol{k}) \\
            H^{\dagger}_{\uparrow\downarrow}(\boldsymbol{k}) &
            H_{\downarrow\downarrow}(\boldsymbol{k})
        \end{array}
        \right),\\
    &= \left( 
        \begin{array}{cccc}
            H_{1\uparrow,1\uparrow}(\boldsymbol{k}) &
            H_{1\uparrow,2\uparrow}(\boldsymbol{k}) &
            H_{1\uparrow,1\downarrow}(\boldsymbol{k}) &
            H_{1\uparrow,2\downarrow}(\boldsymbol{k}) \\
             H^*_{1\uparrow,2\uparrow}(\boldsymbol{k}) &
            H_{2\uparrow,2\uparrow}(\boldsymbol{k}) &
            H_{2\uparrow,1\downarrow}(\boldsymbol{k}) &
            H_{2\uparrow,2\downarrow}(\boldsymbol{k}) \\
              H^*_{1\uparrow,1\downarrow}(\boldsymbol{k}) &
            H^*_{2\uparrow,1\downarrow}(\boldsymbol{k}) &
            H_{1\downarrow,1\downarrow}(\boldsymbol{k}) &
            H_{1\downarrow,2\downarrow}(\boldsymbol{k}) \\
             H^*_{1\uparrow,2\downarrow}(\boldsymbol{k}) &
            H^*_{2\uparrow,2\downarrow}(\boldsymbol{k}) &
            H^*_{1\downarrow,2\downarrow}(\boldsymbol{k}) &
            H_{2\downarrow,2\downarrow}(\boldsymbol{k}) 
        \end{array}
    \right),
\end{aligned}
\end{equation}
where we have imposed hermiticity. 

Let us list the action on the Bloch orbital basis of a few symmetries that we use below, 
\begin{equation}
\label{eq_symmetries}
    \begin{aligned}
        ^{C_{2z}} \vert \boldsymbol{\varphi}, \boldsymbol{k} \rangle &= \vert \boldsymbol{\varphi}, C_{2z}\boldsymbol{k} \rangle \left(-\mathrm{i}\sigma_z \otimes \mathbb{1} \right),\\
        ^{I} \vert \boldsymbol{\varphi}, \boldsymbol{k} \rangle &= \vert \boldsymbol{\varphi}, -\boldsymbol{k} \rangle \left(\mathbb{1} \otimes \mathbb{1} \right) ,\\
        ^{m_z} \vert \boldsymbol{\varphi}, \boldsymbol{k} \rangle &= \vert \boldsymbol{\varphi}, m_z\boldsymbol{k} \rangle \left(-\mathrm{i}\sigma_z \otimes \mathbb{1} \right),\\
        ^{\mathcal{T}} \vert \boldsymbol{\varphi}, \boldsymbol{k} \rangle &= \vert \boldsymbol{\varphi}, -\boldsymbol{k} \rangle \left(-\mathrm{i}\sigma_y \otimes \mathbb{1} \right) \mathcal{K},\\
        ^{C_{2z}\mathcal{T}} \vert \boldsymbol{\varphi}, \boldsymbol{k} \rangle &= \vert \boldsymbol{\varphi}, m_z \boldsymbol{k} \rangle \left(\mathrm{i}\sigma_x \otimes \mathbb{1} \right) \mathcal{K},\\
        ^{I\mathcal{T}} \vert \boldsymbol{\varphi}, \boldsymbol{k} \rangle &= \vert \boldsymbol{\varphi}, \boldsymbol{k} \rangle \left(-\mathrm{i}\sigma_y \otimes \mathbb{1} \right) \mathcal{K},
    \end{aligned}
\end{equation}
where $C_{2z}$ is the $\pi$ rotation about the $\hat{z}$ axis that is perpendicular to the basal plane containing the two-dimensional system, $I$ is inversion, $m_z = C_{2z}I= IC_{2z}$ is the basal mirror, $\mathcal{T}$ is time reversal, and $\mathcal{K}$ is complex conjugation. 

We now consider a system that is symmetric under $C_{2z}\mathcal{T}$ only, {\it i.e.} it must satisfy the constraint
\begin{equation}
    \left(\sigma_x \otimes \mathbb{1} \right) H^*(m_z \boldsymbol{k}) \left(\sigma_x \otimes \mathbb{1} \right) = H(\boldsymbol{k}).
\end{equation}
In 2D systems, the momenta belong to the $m_z$-invariant Brillouin zone, {\it i.e.}~$m_z \boldsymbol{k} = \boldsymbol{k}$. As a consequence the blocks $H_{\sigma\sigma'}(\boldsymbol{k})$ that compose $H(\boldsymbol{k})$ must be of the form
\begin{equation}
\begin{aligned}
    H_{\uparrow\uparrow} &= H_1 \sigma_x + H_2 \sigma_y + H_3 \sigma_z + H_4 \mathbb{1} ,\\
    H_{\downarrow\downarrow} &= H_1 \sigma_x - H_2 \sigma_y + H_3 \sigma_z + H_4 \mathbb{1} ,\\
    H_{\uparrow\downarrow} &= \left(\begin{array}{cc} 
        H_5+\mathrm{i} H_6 & H_7+\mathrm{i} H_8 \\
        H_7+\mathrm{i} H_8 & H_9+\mathrm{i} H_{10}
    \end{array}\right),
\end{aligned}
\end{equation}
where
\begin{equation}
    \begin{array}{rclcrcl}
        H_1 &=& \Re H_{1\uparrow,2\uparrow} ,& H_6 &=& \Im H_{1\uparrow,1\downarrow} ,\\
        H_2 &=& -\Im H_{1\uparrow,2\uparrow} ,&H_7 &=& \Re H_{1\uparrow,2\downarrow}  ,\\
        H_3 &=&  (H_{1\uparrow,1\uparrow}-H_{2\uparrow,2\uparrow})/2 ,&H_8 &=& \Im H_{1\uparrow,2\downarrow} ,\\
        H_4 &=& (H_{1\uparrow,1\uparrow}+H_{2\uparrow,2\uparrow})/2 ,& H_9 &=& \Re H_{2\uparrow,2\downarrow} ,\\
        H_5 &=& \Re H_{1\uparrow,1\downarrow} ,&H_{10} &=& \Im H_{2\uparrow,2\downarrow} ,
    \end{array}
\end{equation}

From $[C_{2z}\mathcal{T}]^2 = +1$ follows that there exists a basis in which $C_{2z}\mathcal{T}$ is represented by $\mathcal{K}$ \cite{Bouhon2019c}. This basis is here given by 
\begin{equation}
\label{eq_real_basis}
\begin{aligned}
    \vert \widetilde{\boldsymbol{\varphi}} , \boldsymbol{k} \rangle &= 
    \vert \boldsymbol{\varphi} , \boldsymbol{k} \rangle \cdot V^{\dagger},\\
    V &= \sqrt{\sigma_x\otimes \mathbb{1}} \cdot 
    \dfrac{1}{2}\left(
        \begin{array}{rrrr}
        -1 & \mathrm{i} & 1 & -\mathrm{i} \\ 
        \mathrm{i} & -1 & \mathrm{i} & -1\\  -\mathrm{i} & 1 & \mathrm{i} & -1 \\   
        -1 & \mathrm{i} & -1 & \mathrm{i} 
        \end{array}
    \right), 
\end{aligned}
\end{equation}
for which
\begin{equation}
\begin{aligned}
    ^{C_{2z}\mathcal{T}} \vert \widetilde{\boldsymbol{\varphi}}, \boldsymbol{k} \rangle &= ^{C_{2z}\mathcal{T}}\vert \boldsymbol{\varphi}, \boldsymbol{k} \rangle
    \cdot V^{\dagger},\\
    &= \vert \boldsymbol{\varphi}, \boldsymbol{k} \rangle
    \cdot
    \left(\mathrm{i}\sigma_x \otimes \mathbb{1} \right) V^{T} \mathcal{K},\\
    &= \vert \widetilde{\boldsymbol{\varphi}}, \boldsymbol{k} \rangle
    \cdot V
    \left(\mathrm{i}\sigma_x \otimes \mathbb{1} \right)  V^{T}\mathcal{K},\\
    &= \vert \widetilde{\boldsymbol{\varphi}}, \boldsymbol{k} \rangle \mathcal{K}.
\end{aligned}
\end{equation}
Rotating the Hamiltonian in the new basis, we define 
\begin{equation}
    \widetilde{H}(\boldsymbol{k}) = V\cdot  H(\boldsymbol{k}) \cdot V^{\dagger} ,
\end{equation}
that now must satisfy $\widetilde{H}^*(\boldsymbol{k}) =\widetilde{H}(\boldsymbol{k})$ as a consequence of $C_{2z}\mathcal{T}$ symmetry, {\it i.e.} $\widetilde{H}(\boldsymbol{k})$ is real and symmetric. 

In the ``real'' basis, the Hamiltonian thus has the generic form
\begin{equation}
\label{eq_real_H}
    \widetilde{H} = \left(\begin{array}{cccc}
        h_{11} & h_{12} & h_{13} & h_{14} \\
        h_{12} & h_{22} & h_{23} & h_{24} \\
        h_{13} & h_{23} & h_{33} & h_{34} \\
        h_{14} & h_{24} & h_{34} & h_{44} 
    \end{array}\right),
\end{equation}
where all elements $h_{ij}$ are real and given by
\begin{equation}
    \begin{array}{rcl}
        h_{11} &=& H_3+H_4-H_5 , \\
        h_{22} &=& H_3+H_4+H_5 , \\
        h_{33} &=& -H_3+H_4-H_9 , \\
        h_{44} &=& -H_3+H_4+H_9 , \\
        h_{12} &=& -H_6 , \\
        h_{13} &=& -H_1+H_7 , \\
        h_{14} &=& H_2-H_8 , \\
        h_{23} &=& H_2+H_8 , \\
        h_{24} &=& H_1+H_7 , \\
        h_{34} &=& H_{10} .
    \end{array}
\end{equation}

\section{Mirror Chern number and Euler class}\label{sec:mz}

Let us assume that the system satisfies the basal mirror symmetry $m_z$ as well. Then, it must also have $ -m_z C_{2z}\mathcal{T}  = -C_{2z}^2 I \mathcal{T} = I\mathcal{T} $ symmetry. The system thus has the symmetries of the magnetic point group $2'/m = \{E,C_{2z}\mathcal{T},m_z,I\mathcal{T}\}$. It readily follows that the off-diagonal blocks $H_{\uparrow\downarrow} = H_{\downarrow\uparrow}$ must vanish, {\it i.e.}~$H_5 = H_6=H_7=H_8=H_9=H_{10} = 0$. The $m_z$-invariant Hamiltonian in the spinor basis then reads
\begin{equation}
    H = H_{\uparrow\uparrow} \oplus H_{\downarrow\downarrow}.
\end{equation}
Since $^{I\mathcal{T}}\mathcal{H} = \mathcal{H}$ with $[I\mathcal{T}]^2 = -1$, the bands must be twofold-degenerate at all momenta, namely the bands are Kramers degenerate. The eigenvalues are indeed readily found to be $\{E_o,E_o,E_u,E_u\}$, with
\begin{equation}
\begin{aligned}
    E_{o} &= H_4 - \epsilon,~E_{u} = H_4 + \epsilon,\\
    \epsilon &= \sqrt{H^2_1+H^2_2+H^2_3}.
\end{aligned}
\end{equation}
We note that $H_4$ can be chosen arbitrarily without affecting the symmetry and the topology, we thus set $H_4 = 0$ without loss of generality. The eigenvectors are 
\begin{equation}
\begin{aligned}
    v^{(\uparrow)}_{o} = 
    \left(\begin{array}{c} 
        \sqrt{1 - r^2} \\
        - r \,\mathrm{e}^{\mathrm{i} \rho} \\
        0 \\
        0
    \end{array}\right),&\;
    v^{(\uparrow)}_{u} = \left(\begin{array}{c} 
        r \\
        \sqrt{1 - r^2} \,\mathrm{e}^{\mathrm{i} \rho} \\
        0 \\
        0
    \end{array}\right),\\
    v^{(\downarrow)}_{o} = 
    \left(\begin{array}{c} 
        0 \\
        0\\
        \sqrt{1 - r^2} \\
        - r \,\mathrm{e}^{-\mathrm{i} \rho} 
    \end{array}\right),&\;
    v^{(\downarrow)}_{u} = \left(\begin{array}{c} 
        0 \\
        0 \\
        r \\
        \sqrt{1 - r^2} \,\mathrm{e}^{-\mathrm{i} \rho} 
    \end{array}\right),
\end{aligned}
\end{equation}
with 
\begin{equation}
    \rho = \mathrm{Arg}\left\{H_1 + i H_2\right\},\quad r^2  = \dfrac{\epsilon + H_3}{2\epsilon}.
\end{equation}

The topology can now be directly assessed from a single spin sector, say from $H_{\uparrow\uparrow}$. Imposing the condition of a band gap, {\it i.e.}~$\epsilon > 0$, we define the unit vector 
\begin{equation}
\label{eq_unitvec}
    \boldsymbol{n}_{\uparrow} = \dfrac{1}{\epsilon} (H_1,H_2,H_3) ,
\end{equation}
in terms of which we obtain the skyrmion number $W[\boldsymbol{n}_{\uparrow}]$ Eq.~(\ref{eq_Skyrmion}).
Characterizing the $H_{\downarrow\downarrow}$ with the unit vector 
\begin{equation}
    \boldsymbol{n}_{\downarrow} =  \dfrac{1}{\epsilon}(H_1,-H_2,H_3),    
\end{equation}
we have
\begin{equation}
    W[\boldsymbol{n}_{\downarrow}] = -W[\boldsymbol{n}_{\uparrow}].
\end{equation}

It can be checked that $W[\boldsymbol{n}_{\uparrow}]$ directly gives the Chern number computed through the surface integral of the Berry curvature for the occupied eigenvector in the $\uparrow$-spin sector, {\it i.e.}~defining
\begin{multline}
    F[v^{(\uparrow)}_o] = -\mathrm{i}\left[(\partial_{k_1}v^{(\uparrow)}_{o})^{\dagger} \cdot (\partial_{k_2}v^{(\uparrow)}_{o}) \right.\\
    \left. -  (\partial_{k_2}v^{(\uparrow)}_{o})^{\dagger} \cdot (\partial_{k_1}v^{(\uparrow)}_{0})
    \right]\;,
\end{multline}
we have
\begin{equation}
    C^{(\uparrow)} = \dfrac{1}{2\pi} \int d^2 \boldsymbol{k} ~F[v^{(\uparrow)}_{o}] = W[\boldsymbol{n}_{\uparrow}],
\end{equation}
and similarly
\begin{equation}
    C^{(\downarrow)} = \dfrac{1}{2\pi} \int d^2 \boldsymbol{k} ~F[v^{(\downarrow)}_{o}] = W[\boldsymbol{n}_{\downarrow}] = -C^{(\uparrow)} .
\end{equation}

Furthermore, since the $m_z$ operator is diagonal in the orbital-spinor basis (see Eq.~(\ref{eq_symmetries})), we readily find the Chern number in the $(-\mathrm{i})$-mirror eigenvalue sector, {\it i.e.}
\begin{equation}
\label{eq_Chern_mirror}
    C^{(-\mathrm{i})} = C^{(\uparrow)} = W[\boldsymbol{n}_{\uparrow}],
\end{equation}
called the mirror Chern number. Similarly, the mirror Chern number of the other spin (mirror) sector is
\begin{equation}
    C^{(\mathrm{i})} = C^{(\downarrow)} = W[\boldsymbol{n}_{\downarrow}] = - W[\boldsymbol{n}_{\uparrow}].
\end{equation}

Moving to the real basis, we now show that there is a one-to-one correspondence between the mirror Chern number of one occupied mirror-polarized band and the signed Euler class of the occupied two-band subspace. First, let us write the $m_z$-invariant Hamiltonian in its real symmetric form,
\begin{equation}
\label{eq_Hreal_inter}
\begin{aligned}
    \widetilde{H} &= 
    \left(\begin{array}{cccc}
        H_3 & 0 & -H_1 & H_2 \\
        0 & H_3 & H_2 & H_1 \\
        -H_1 & H_2 & -H_3 & 0 \\
        H_2 & H_1 & 0 & -H_3 
    \end{array} 
    \right),\\
   &= - H_1 (\sigma_{x}\otimes\sigma_{z} ) + H_2 (\sigma_{x}\otimes\sigma_{x} ) + H_3 (\sigma_{z}\otimes\mathbb{1} ).
\end{aligned}
\end{equation}
Assuming again the gap condition, {\it i.e.}~$\epsilon^2 = H^2_1 + H^2_2+H^2_3>0$, without loss of generality we can deform the Hamiltonian as $H_i \rightarrow n_{\uparrow,i}$ for $i=1,2,3$. 

We now derive the direct relation between the mirror Chern number and the Euler class of the system via the Pl{\"u}cker embedding. First, we parametrize the flattened Hamiltonian Eq.\;(\ref{eq_Hreal_inter}) through 
\begin{equation}
    \begin{aligned}
        H_1/\epsilon &= n_{\uparrow,1} =  \cos \phi \sin\theta, \\
        H_2/\epsilon &= n_{\uparrow,2}= \sin\phi\sin\theta, \\
        H_3/\epsilon &= n_{\uparrow,3} = \cos \theta .
    \end{aligned}
\end{equation}
The eigenvalues are then $\{-1,-1,1,1\}$, and the two real eigenvectors of the occupied bands are given by
\begin{equation}
    \begin{aligned}
        u_1 (\phi,\theta) &= \sqrt{\cos(\theta/2)^2}\left(\begin{array}{c}
            \sin\phi\tan(\theta/2) \\
            \cos\phi\tan(\theta/2) \\
            0 \\ -1
        \end{array}
        \right),\\
        u_2 (\phi,\theta) &= \sqrt{\cos(\theta/2)^2}\left(\begin{array}{c} 
            -\cos\phi\tan(\theta/2)\\
            \sin\phi\tan(\theta/2)\\
            -1 \\ 0 \end{array}
        \right).
    \end{aligned}
\end{equation}

Then, the wedge product of the two occupied bands (Pl{\"u}cker embedding \cite{BouhonGEO2020}) gives
\begin{equation}
\label{eq_winding_Euler}
    \begin{aligned}
    \left(\begin{array}{c}
        u_1^3 u_2^2 - u_1^2 u_2^3 \\
        u_1^3 u_2^1 - u_1^1 u_2^3 \\
        u_1^1 u_2^2 - u_1^2 u_2^1 \\
        u_1^4 u_2^1 - u_1^1 u_2^4 \\
        u_1^2 u_2^4 - u_1^4 u_2^2 \\
        u_1^3 u_2^4 - u_1^4 u_2^3 
    \end{array}\right) \cdot&
    \left(\begin{array}{cc}
        \mathbb{1}_3 & \mathbb{1}_3 \\
        \mathbb{1}_3 & -\mathbb{1}_3
    \end{array}\right) = 
    \left(\begin{array}{c}
        \cos\phi\sin\theta \\
        \sin\phi \sin\theta \\
        -\cos \theta \\
        0\\
        0\\
        1
    \end{array}
    \right)\\
    &= \left(
        n_{1,\uparrow}, n_{2,\uparrow}, -n_{3,\uparrow}, 0, 0, 1 \right)^T.
    \end{aligned}
\end{equation}
The Euler class is defined as the winding of the wedge product. The above relation thus explicitly shows that the Euler class of the two occupied bands is readily given by the (oriented) number of times $\boldsymbol{n}_{\uparrow}(\phi(k_x,k_y),\theta(k_x,k_y))$ wraps the sphere when $(k_x,k_y)$ covers the Brillouin zone one time \cite{BouhonGEO2020}. 
The proof is completed by noting that the Hamiltonian Eq.\;(\ref{eq_Hreal_inter}), after flattening the eigenvalues, is defined in terms of Eq.\;(\ref{eq_H_general}) by
\begin{equation}
    \widetilde{H}[H_i\rightarrow n_{\uparrow,i}] = \widetilde{H}_E[(n_{\uparrow,1},n_{\uparrow,2},-n_{\uparrow,3}),(0,1,0)].
\end{equation}
The Euler class of the system are then determined from the Skyrmion numbers 
\begin{equation}
\begin{aligned}
    q &= W[(n_{\uparrow,1},n_{\uparrow,2},-n_{\uparrow,3})] = -W[\boldsymbol{n}_{\uparrow}], \\
    q'&=0,
\end{aligned}
\end{equation}
via Eq.\;(\ref{eq_winding}), to be $(E_I,E_{II})=(q,q)$. Then with
Eq.~(\ref{eq_Chern_mirror}), we obtain
\begin{equation}
\label{eq_Euler_chern_mirror}
    E_{I} = E_{II} = - C^{(-\mathrm{i})} \in \mathbb{Z}.
\end{equation}

A few comments are needed here. In general ({\it i.e.}~without mirror symmetry), the sign of the Euler class is not uniquely defined because, as noted above, Hamiltonians only define \textit{orientable} vector bundles (instead of \textit{oriented} vector bundles) as a consequence of the gauge freedom $u_i \rightarrow \pm u_i$ for every eigenvector, which allows flipping the sign of the wedge product between the two occupied eigenvectors, {\it i.e.}~$u_1 \wedge u_2 \rightarrow - u_1 \wedge u_2$. This has the consequence that there exist adiabatic transformations of the Hamiltonian that flip the sign of the pair of Euler classes, leading to the topological equivalence $(E_I,E_{II})\sim (-E_I,-E_{II})$ \cite{BouhonGEO2020}. Under the constraint of the basal mirror symmetry $m_z$ though, one can associate a signed winding number to a fixed mirror eigenvalue sector. 

Let us write the representation of $m_z$ in the basis of real eigenvectors, {\it i.e.}   
\begin{equation}
\begin{aligned}
    \left(\begin{array}{cc}
         u_1^T  \\
         u_2^T 
    \end{array}\right) \cdot \widetilde{U}_{m_z} \cdot \left(u_1~u_2\right) = \left(\begin{array}{cc} 
        0 & 1\\
        -1  & 0
    \end{array}\right),
\end{aligned}
\end{equation}
where 
\begin{equation}
\label{eq_mz_real_basis}
    \widetilde{U}_{m_z} = V^{\dagger}\cdot (-\mathrm{i} \sigma_z \otimes \mathbb{1}) \cdot V = \sigma_z \otimes \mathrm{i} \sigma_y    
\end{equation} 
is the representation of $m_z$ in the ``real'' Bloch orbital basis Eq.~(\ref{eq_real_basis}). The eigenbasis of $m_z$ is thus given through the complexification (see the Chern basis in the main text)
\begin{equation}
\begin{aligned}
    v_+ &= (u_1+\mathrm{i} u_2)\sqrt{2},\\ 
    v_- &= (u_1-\mathrm{i} u_2)/\sqrt{2},
\end{aligned}
\end{equation} 
{\it i.e.}
\begin{equation}
\label{eq_mirror_basis}
\begin{aligned}
    \left(\begin{array}{cc}
         v_+^{\dagger}  \\
         v_-^{\dagger} 
    \end{array}\right) \cdot \widetilde{U}_{m_z} \cdot \left(v_+~v_-\right) = \left(\begin{array}{cc} 
        -\mathrm{i} & 0\\
        0  & \mathrm{i}
    \end{array}\right) = -\mathrm{i} \sigma_z,
\end{aligned}
\end{equation}
which is now diagonal, such that $v_{\pm}$ are eigenvectors of $m_z$ with the eigenvalues $\mp\mathrm{i}$. (Note that we actually recover the action of $m_z$ on the spinor basis given in Eq.~(\ref{eq_symmetries}).) It is now transparent from Eq.~(\ref{eq_mirror_basis}) that the relative sign between $u_1$ and $u_2$, and thus the sign of the Euler class (fixed by the wedge product $u_1 \wedge u_2$), is fixed by the mirror symmetry, since the gauge transformation $u_1\wedge u_2 \rightarrow -u_1\wedge u_2$ implies $v_{\pm} \rightarrow v_{\mp}$ which is forbidden under the constraint of a fixed mirror-eigenvalue sector. Note that the gauge transformation $(u_1,u_2) \rightarrow (-u_1,-u_2)$ is allowed since it doesn't change the fixed mirror-eigenvalue sector, nor does it change the signed Euler class.  

We emphasise that $\{v_1,v_2\}$ are still eigenvectors of the Hamiltonian since the energy eigenvalues for $u_1$ and $u_2$ are degenerate. Furthermore, we readily recover the $(-\mathrm{i})$- and $(\mathrm{i})$-mirror Chern numbers as the Chern numbers of $v_1$ and $v_2$, respectively. It is now apparent that the winding associated to a nontrivial Euler class in Eq.~(\ref{eq_winding_Euler}), directly implies the winding associated to the mirror Chern numbers, according to Eq.~(\ref{eq_Chern_mirror}). We conclude that by imposing that $u_1 + \mathrm{i} u_2$ has the mirror eigenvalue $-\mathrm{i}$, there is a one-to-one correspondence between the Euler class Eq.~(\ref{eq_winding_Euler}) and the mirror Chern number Eq.~(\ref{eq_Chern_mirror}), leading to Eq.~(\ref{eq_Euler_chern_mirror}).

\subsection{All degenerate balanced phases are mirror-symmetric}\label{ap_thm_deg_balanced_mirror}

Importantly, the above reasoning for balanced Euler insulating phases can be reversed. Namely, for every (orientable) balanced ($E_I=E_{II}$) topological phase with only the $C_{2z}\mathcal{T}$ symmetry, whenever we impose the two-by-two degeneracy of the bands, there must be an effective basal mirror symmetry $m_z$ (spinful with $m_z^2=-1$), leading to the effective $I\mathcal{T}$ symmetry with $[I\mathcal{T}]^2 = -1$. In other words, the degeneracy of the bands is always associated with a symmetry of the Hamiltonian and no fine-tuning is needed.  

We prove this by going back to the general geometric form from which all our tight-binding Hamiltonian are derived, Eq.~(\ref{eq_H_general}). First of all, all Hamiltonian belonging to $\mathrm{Gr}^{\mathbb{R}}_{2,4}$ can be adiabatically mapped to a twofold-degenerate Hamiltonian. By construction the representative of each (orientable) homotopy class $\widetilde{H}[\boldsymbol{n},\boldsymbol{n}']$ is twofold degenerate. Without loss of generality, the balanced phases are obtained by fixing $\boldsymbol{n}'$ to be non-winding, {\it i.e.}~$q'= W[\boldsymbol{n}']  = 0$. Note that we can alternatively fix $\boldsymbol{n}$ to a constant and let $\boldsymbol{n}'$ wind instead, \ie the transformation $(q,q'=0) \rightarrow (0,q'=q)$, which induces the flip of one Euler class. That is from Eq.\;(\ref{eq_winding}) 
\begin{equation}
    (E_{I},E_{II}) \longrightarrow (E'_{I},E'_{II}) = (-E_{I},E_{II}).
\end{equation}
Since we do not observe any qualitative difference in the Hofstadter spectrum between the phases $(E_{I},E_{II})$ and $\pm (-E_{I},E_{II})$, we have shown results for $E_I,E_{II} \geq 0$ only.

Let us fix $\boldsymbol{n} = (H_1,H_2,-H_3)/\epsilon$ and $\boldsymbol{n}' = \bar{\boldsymbol{n}} = (0,1,0) $ in Eq.\;(\ref{eq_H_general}), which leads to the Hamiltonian Eq.\;(\ref{eq_Hreal_inter}). We find that it is mirror symmetric with
\begin{equation}
\label{eq_Hgr_mirror}
    \widetilde{U}_{m_z} \cdot \widetilde{H}[\boldsymbol{n},\bar{\boldsymbol{n}}] \cdot \widetilde{U}_{m_z}^{T} = \widetilde{H}[\boldsymbol{n},\bar{\boldsymbol{n}}],
\end{equation}
where $\widetilde{U}_{m_z}$ is defined in Eq.~(\ref{eq_mz_real_basis}). Comparing $\widetilde{H}[\boldsymbol{n},\boldsymbol{n}']$ with two different fixed unit vectors $\boldsymbol{n}'$, {\it i.e.} in one case $\boldsymbol{n}'=\bar{\boldsymbol{n}}$ and in the other case $\boldsymbol{n}' = \boldsymbol{n}_1 \neq \bar{\boldsymbol{n}}$, we find
\begin{multline}
\label{eq_H_transfer}
    \widetilde{H}[\boldsymbol{n},\bar{\boldsymbol{n}}] =\\
    \Delta R[\boldsymbol{n},\bar{\boldsymbol{n}},\boldsymbol{n}_1] \cdot \widetilde{H}[\boldsymbol{n},\boldsymbol{n}_1] \cdot \Delta R[\boldsymbol{n},\bar{\boldsymbol{n}},\boldsymbol{n}_1] ^T,
\end{multline}
with 
\begin{equation}
    \Delta R[\boldsymbol{n},\bar{\boldsymbol{n}},\boldsymbol{n}_1]  = R[\boldsymbol{n},\bar{\boldsymbol{n}}]\cdot R[\boldsymbol{n},\boldsymbol{n}_1]^T .
\end{equation}
Substituting Eq.~(\ref{eq_H_transfer}) in Eq.~(\ref{eq_Hgr_mirror}), we then obtain
\begin{equation}
\label{eq_mirror_H_deformed}
     \Delta\widetilde{U}_{m_z}(\boldsymbol{n}_1) \cdot \widetilde{H}[\boldsymbol{n},\boldsymbol{n}_1] \cdot \Delta\widetilde{U}_{m_z}(\boldsymbol{n}_1)^{T} = \\ \widetilde{H}[\boldsymbol{n},\boldsymbol{n}_1],
\end{equation}
with
\begin{equation}
\label{eq_deformed_mirror}
\begin{aligned}
    \Delta\widetilde{U}_{m_z}(\boldsymbol{n}_1) & = \Delta R[\boldsymbol{n},\bar{\boldsymbol{n}},\boldsymbol{n}_1] ^T \cdot \widetilde{U}_{m_z} \cdot \Delta R[\boldsymbol{n},\bar{\boldsymbol{n}},\boldsymbol{n}_1] ,
\end{aligned}
\end{equation}
{\it i.e.}~the deformed Hamiltonian $\widetilde{H}[\boldsymbol{n},\bar{\boldsymbol{n}}]\rightarrow \widetilde{H}[\boldsymbol{n},\boldsymbol{n}_1]$ is still mirror symmetric, with a constant mirror operator $\Delta\widetilde{U}_{m_z}(\boldsymbol{n}_1)$. 

Interestingly, we can consider more general adiabatic transformations for which $\boldsymbol{n}'$ is non-constant but still non-winding, {\it i.e.}~$\boldsymbol{n}' = \boldsymbol{n}'(\boldsymbol{k}) $ with $W[\boldsymbol{n}'(\boldsymbol{k})]=0$. In that case, the ``mirror'' symmetry operator is non-constant, with a nontrivial action of the ``mirror'' symmetry on the momentum. We will explore such phases elsewhere. 

Since our homotopy representative Hamiltonian of the balanced Euler insulating phases covers all the balanced homotopy classes (with $q'=0$ and $q \in \mathbb{Z}$), we conclude with the following statement: \textit{Every (two-dimensional, four-band at half-filling, orientable) balanced Euler insulating phase is (spinful) mirror-symmetric with respect to the basal plane if and only if the energy eigenvalues are twofold-degenerate.}

\subsection{Degenerate imbalanced phases}

Contrary to the balanced case, the imbalanced Euler insulating phases ($\vert E_I\vert \neq \vert E_{II}\vert$) with twofold-degenerate energy eigenvalues are never compatible with an effective mirror symmetry $m_z$, and thus there is no Kramers degeneracy. As a consequence, the degeneracy of the bands for these phases always requires fine-tuning. 

Let us prove this. The imbalanced condition imposes that $q,q' \neq 0$, {\it i.e.}~the two-unit vectors $\boldsymbol{n}$ and $\boldsymbol{n}'$ both wind. We simply define the imbalanced Hamiltonian from the balanced one through Eq.~(\ref{eq_H_transfer}), where we substitute the constant unit vector $\boldsymbol{n}_1$ to the winding one $\boldsymbol{n}'$. As a consequence, the degenerate imbalanced Hamiltonian satisfies Eq.~(\ref{eq_mirror_H_deformed}) but now with a mirror symmetry operator $\Delta\widetilde{U}_{m_z}(\boldsymbol{n}')$ in Eq.~(\ref{eq_deformed_mirror}) that winds. Therefore, the condition Eq.~(\ref{eq_mirror_H_deformed}) cannot be interpreted as the symmetry of one fixed homotopy class. Since the degeneracy of the bands in one imbalanced homotopy class is never associated with a global symmetry of the Hamiltonian, it is accidental and can only be realized through fine-tuning. 

It can be verified with the Mathematica notebook of Ref.~\cite{abouhon_EulerClassTightBinding} that the degeneracy of the bands of imbalanced phases is never exact whenever we truncate the Fourier expansion of Eq.~(\ref{eq_Fourier}). However, since the hopping parameters decay rapidly, similarly to the flat-band limit, we obtain degenerate bands in a good numerical approximation, see {\it e.g.}~Fig.~\ref{fig:BS_flat_band}(b) showing the band structure of the phase $(E_I,E_{II}) = (1,3)$ obtained for $K=12$ where both degeneracy and flatness have been imposed.

\section{Hidden symmetry of the dispersive balanced phases and comparison with the QHS model} \label{sec:hidden}

The non-degenerate phases break the mirror symmetry $m_z$. The constraint of the $C_{2z}\mathcal{T}$ symmetry alone allows all the terms of the real symmetric Hamiltonian in Eq.~(\ref{eq_real_H}) to be nonzero. Form the systematic probe of all allowed (adiabatic) perturbations of the model $\widetilde{H}_{\mathrm{bal}}[\boldsymbol{n}] = \widetilde{H}[\boldsymbol{n},(0,1,0)]$ given by Eq.~(\ref{eq_H_general}) (see also the main text), we have found that the gapping of the Hofstadter spectrum at half-filling only happens when $ h_{14} \neq h_{23}$. Setting $h_{23}=h+\delta$ and $h_{14}=h-\delta$, a general (real symmetric) balanced Hamiltonian then takes the form
\begin{equation}
    \widetilde{H}_{\mathrm{bal}} = 
    \widetilde{H}_{\mathrm{bal}}[\delta=0] + \delta (\sigma_y \otimes \sigma_y),
\end{equation}
with $\delta = (h_{23} - h_{14})/2$.

We call the condition $\delta=0$ a hidden symmetry of the balanced Hamiltonian at finite flux. In other words, every tight-binding Hamiltonian that is of the form $\widetilde{H}_{\mathrm{bal}}[\delta=0]$ satisfies the hidden symmetry and exhibits a gapless Hofstadter spectrum. On the contrary, any model with $\vert \delta\vert > 0$ has a gapped Hofstadter spectrum.

\subsubsection{Comparison with the QSH model}

In Ref.~\cite{Herzog-Arbeitman2020a} the authors have considered the BHZ model of the Quantum Spin Hall phase (QSH) and its generalization when only the $C_{2z}\mathcal{T}$ symmetry is preserved and restricting to balanced phases, which they call $H_{QSH}'''$. For exhaustiveness, we give here the mapping from $H_{QSH}'''$ (which is not in its real form) to our models in Eq.~(\ref{eq_real_H}).

We first write $H_{QSH}'''$ in its generic form, {\it i.e.}
\begin{multline}
    H_{QSH}''' = a_0 (\mathbb{1}\otimes \mathbb{1}) + a_1 (\mathbb{1}\otimes\sigma_z) +
    a_2 (\sigma_z\otimes\sigma_x) +\\
    a_3 (\sigma_z\otimes\sigma_x) +
    a_4 (\sigma_y\otimes\sigma_z) +
    a_5 (\sigma_x\otimes\mathbb{1}) +
    a'_5 (\sigma_y\otimes\mathbb{1}) +\\
    a_6 (\sigma_x\otimes\sigma_y) +
    a'_6 (\sigma_y\otimes\sigma_y) +
    a''_6 (\sigma_x\otimes\sigma_z) ,
\end{multline}
where we have added the term $a_4 (\sigma_y\otimes\sigma_z)$ which is allowed by $C_{2z}\mathcal{T}$ but not present in the model of Ref.~\cite{Herzog-Arbeitman2020a}. (We note that we are not concerned here with the specific expressions of the terms of $H_{QSH}'''$ given in Ref.~\cite{Herzog-Arbeitman2020a} which realizes the phase $E_{I}=E_{II}=1$.) 

We now perform a change of basis that brings $H_{QSH}'''$ in its real form in order to compare it with our models. We define
\begin{equation}
    \widetilde{H}_{QSH}''' = P\cdot V\cdot H_{QSH}''' \cdot V^{\dagger} \cdot P,
\end{equation}
that is real and symmetric, with 
\begin{equation}
        V = \sqrt{\sigma_x \otimes \sigma_z} ,\quad
        P = \left(\begin{array}{cccc}
            1 & 0 & 0 & 0 \\
            0 & 0 & 1 & 0 \\
            0 & 1 & 0 & 0 \\
            0 & 0 & 0 & 1 
        \end{array}\right).
\end{equation}
Writing it explicitly, we have 
\begin{equation}
    \widetilde{H}_{QSH}''' 
    = \left(\begin{array}{cccc}
        h_{11} & h_{12} & h_{13} & h_{14} \\
        h_{12} & h_{22} & h_{23} & h_{24} \\
        h_{13} & h_{23} & h_{33} & h_{34} \\
        h_{14} & h_{24} & h_{34} & h_{44} 
    \end{array}\right),
\end{equation}
with 
\begin{equation}
\begin{aligned}
    h_{11} & = a_0+a_1+a_4+a_5' \\
    h_{22} & = a_0+a_1-a_4-a_5' \\
    h_{33} & = a_0-a_1+a_4-a_5' \\
    h_{44} & = a_0-a_1-a_4+a_5' \\
    h_{12} & = a_5+a_6'' \\
    h_{13} & = a_2 - a_6 \\
    h_{14} & = -a_3-a_6' \\
    h_{23} & = -a_3 + a'_6,\\
    h_{24} & = -a_2-a_6,\\
    h_{34} & = a_5 -a_6''.
\end{aligned}
\end{equation}

In agreement with our finding of the hidden symmetry, we have verified that only the nonzero term $a_6' (\sigma_y\otimes\sigma_y)$ of $H_{QSH}'''$ leads to the gapping of the Hofstadter spectrum. We indeed have 
\begin{equation}
    a_6' = \dfrac{h_{23}-h_{14}}{2}  = \delta,
\end{equation}
which is the term responsible for the hidden symmetry discussed above.

\bibliography{collec2a}

\appendix

\title{\textsf{Supplementary Information for "Landau Levels of the Euler Class Topology"}}

\author{Guan Yifei}\thanks{Contributed equally.  Correspondence to \href{mailto:yifei.guan@epfl.ch}{yifei.guan@epfl.ch} and \href{mailto:adrien.bouhon@su.se}{adrien.bouhon@su.se}.}
\affiliation{Institute of Physics, \'{E}cole Polytechnique F\'{e}d\'{e}rale de Lausanne (EPFL), CH-1015 Lausanne, Switzerland}

\author{Adrien Bouhon}\thanks{Contributed equally.  Correspondence to \href{mailto:yifei.guan@epfl.ch}{yifei.guan@epfl.ch} and \href{mailto:adrien.bouhon@su.se}{adrien.bouhon@su.se}.}
\affiliation{Nordic Institute for Theoretical Physics (NORDITA), Stockholm, Sweden}

\author{Oleg V. Yazyev}
\affiliation{Institute of Physics, \'{E}cole Polytechnique F\'{e}d\'{e}rale de Lausanne (EPFL), CH-1015 Lausanne, Switzerland}
\affiliation{National Centre for Computational Design and Discovery of Novel Materials MARVEL, \'{E}cole Polytechnique F\'{e}d\'{e}rale de Lausanne (EPFL), CH-1015 Lausanne, Switzerland}

\end{document}